\documentclass[a4paper,fleqn,usenatbib]{mnras}

\usepackage{newtxtext,newtxmath}

\usepackage[T1]{fontenc}
\usepackage{ae,aecompl}

\usepackage{graphicx}	
\usepackage{amsmath}	
\usepackage{amssymb}	

\newcommand{\aref}[1]{\hyperref[#1]{Appendix~\ref{#1}}}

\title[SLRs from Galactic-scale star formation]{Short-lived radioisotopes in meteorites from Galactic-scale correlated star formation}

\author[Y. Fujimoto, M. R. Krumholz and S. Tachibana]{
Yusuke Fujimoto,$^{1}$\thanks{E-mail: yusuke.fujimoto@anu.edu.au}
Mark R. Krumholz$^{1}$
and Shogo Tachibana$^{2}$
\\
$^{1}$Research School of Astronomy \& Astrophysics, Australian National University, Canberra, Australian Capital Territory 2611, Australia\\
$^{2}$UTokyo Organization for Planetary and Space Science (UTOPS), The University of Tokyo, 7-3-1 Hongo, Tokyo 113-0033, Japan
}

\date{Accepted XXX. Received YYY; in original form ZZZ}

\pubyear{2018}

\begin{document}
\label{firstpage}
\pagerange{\pageref{firstpage}--\pageref{lastpage}}
\maketitle

\begin{abstract}
Meteoritic evidence shows that the Solar system at birth contained significant quantities of short-lived radioisotopes (SLRs) such as ${}^{60}\mbox{Fe}$ and ${}^{26}\mbox{Al}$ (with half-lives of 2.6 and 0.7 Myr respectively) produced in supernova explosions and in the Wolf-Rayet winds that precede them. Proposed explanations for the high SLR abundance include formation of the Sun in a supernova-triggered collapse or in a giant molecular cloud (GMC) that was massive enough to survive multiple supernovae (SNe) and confine their ejecta. However, the former scenario is possible only if the Sun is a rare outlier among massive stars, while the latter appears to be inconsistent with the observation that $^{26}$Al is distributed with a scale height significantly larger than GMCs. In this paper, we present a high-resolution chemo-hydrodynamical simulation of the entire Milky-Way Galaxy, including stochastic star formation, H \textsc{ii} regions, SNe, and element injection, that allows us to measure for the distribution of $^{60}\mbox{Fe} / {}^{56}\mbox{Fe}$ and $^{26}\mbox{Al} / {}^{27}\mbox{Al}$ ratios over all stars in the Galaxy. We show that the Solar System's abundance ratios are well within the normal range, but that SLRs originate neither from triggering nor from confinement in long-lived clouds as previously conjectured. Instead, we find that SLRs are abundant in newborn stars because star formation is correlated on galactic scales, so that ejecta preferentially enrich atomic gas that will subsequently be accreted onto existing GMCs or will form new ones. Thus new generations of stars preferentially form in patches of the Galaxy contaminated by previous generations of stellar winds and supernovae.
\end{abstract}



\vspace{1in}


\section{Introduction}



Short-lived radioisotopes (SLRs) -- ${}^{10}\mbox{Be}$, ${}^{26}\mbox{Al}$, ${}^{36}\mbox{Cl}$, ${}^{41}\mbox{Ca}$, ${}^{53}\mbox{Mn}$, ${}^{60}\mbox{Fe}$, ${}^{107}\mbox{Pd}$, ${}^{129}\mbox{I}$, ${}^{182}\mbox{Hf}$ and ${}^{244}\mbox{Pu}$ -- are radioactive elements with half-lives ranging from 0.1 Myr to more than 15 Myr that existed in the early Solar system (e.g., \citealt{Adams2010}). They were incorporated into meteorites' primitive components such as calcium-aluminum-rich inclusions (CAIs), which are the oldest solids in the Solar protoplanetary disc, or chondrules, which formed $\sim$ 1 Myr after CAI formation. The radioactive decay of these SLRs fundamentally shaped the thermal history and interior structure of planetesimals in the early Solar system, and thus is of central importance for core-accretion planet formation models. The SLRs, particularly ${}^{26}\mbox{Al}$, were the main heating sources for the earliest planetesimals and planetary embryos from which terrestrial planets formed \citep{GrimmMcSween1993, JohansenEtAl2015}, and are responsible for the differentiation of the parent bodies of magmatic meteorites in the first few Myrs of the Solar system \citep{GreenwoodEtAl2005, ScherstenEtAl2006, SahijpalSoniGupta2007}. The SLRs are, moreover, potential high-precision and high-resolution chronometers for the formation events of our Solar system due to their short half-lives \citep{KitaEtAl2005, KrotEtAl2008, AmelinEtAl2010, BouvierWadhwa2010, ConnellyEtAl2012}.


Detailed analyses of meteorites show that the early Solar system contained significant quantities of SLRs. The presence of ${}^{26}\mbox{Al}$ in the early Solar system was first identified in CAIs from the primitive meteorite Allende in 1976, defining a canonical initial $^{26}\mbox{Al} / {}^{27}\mbox{Al}$ ratio of $\sim 5 \times 10^{-5}$ \citep{LeeEtAl1976, LeeEtAl1977, JacobsenEtAl2008}, far higher than the ratio of $^{26}\mbox{Al} / {}^{27}\mbox{Al}$ in the interstellar medium (ISM) as estimated from continuous galactic nucleosynthesis models \citep{MeyerClayton2000} and $\gamma$-ray observations measuring the in-situ decay of ${}^{26}\mbox{Al}$  \citep{DiehlEtAl2006}.

Compared to $^{26}\mbox{Al} / {}^{27}\mbox{Al}$, the initial ratio of $^{60}\mbox{Fe} / {}^{56}\mbox{Fe}$ is still somewhat uncertain; analyses of bulk samples of different meteorite types produced a low initial ratio of $^{60}\mbox{Fe} / {}^{56}\mbox{Fe}$ $\sim 1.15 \times 10^{-8}$ \citep{TangDauphas2012, TangDauphas2015}, while other studies of chondrules using in situ measurements found higher initial ratio of $^{60}\mbox{Fe} / {}^{56}\mbox{Fe}$ $\sim 5-13 \times 10^{-7}$ than the ISM ratio (e.g., \citealt{MishraGoswami2014}). \citet{TelusEtAl2016} found that the bulk sample estimates were skewed toward low initial $^{60}\mbox{Fe} / {}^{56}\mbox{Fe}$ ratios because of fluid transport of Fe and Ni during aqueous alteration on the parent body and/or during terrestrial weathering, and \citet{TelusEtAl2018} have found the initial ratios of $^{60}\mbox{Fe} / {}^{56}\mbox{Fe}$ as high as $\sim 0.85-5.1 \times 10^{-7}$, although the initial $^{60}\mbox{Fe} / {}^{56}\mbox{Fe}$ value is still a matter of debate. If estimates in the middle or high end of the plausible range prove to be correct, they would imply a $^{60}\mbox{Fe} / {}^{56}\mbox{Fe}$ ratio well above the interstellar average as well.
 
It has been long debated how the early Solar System came to have SLR abundances well above the ISM average. The isotopes $^{26}\mbox{Al}$ and $^{60}\mbox{Fe}$, on which we focus in this paper, are of particular interest because they are synthesised only in the late stages of massive stellar evolution, followed by injection into the ISM by stellar winds and supernovae (SNe) \citep{HussEtAl2009}. Other SLRs (e.g., ${}^{10}\mbox{Be}$, ${}^{36}\mbox{Cl}$ and ${}^{41}\mbox{Ca}$) can be produced in situ by irradiation of the protoplanetary disc by the young Sun \citep{HeymannDziczkaniec1976, ShuShangLee1996, LeeEtAl1998, ShuEtAl2001, GounelleEtAl2006}.\footnote{Small amounts of $^{26}\mbox{Al}$ can also be produced by this mechanism, but much too little to explain the observed $^{26}\mbox{Al} / {}^{27}\mbox{Al}$ ratio \citep{DupratTatischeff2007}.} Explaining the origin site of the $^{26}\mbox{Al}$ and $^{60}\mbox{Fe}$, and how they travelled from this site to the primitive Solar System before decaying, is an outstanding problem.

One possible origin site is asymptotic giant branch (AGB) stars \citep{WasserburgEtAl1994, BussoGallinoWasserburg1999, WasserburgEtAl2006}. However, because AGB stars only provide SLRs at the end of their lives, and because their main-sequence lifetimes are long ($> 1$ Gyr), the probability of a chance encounter between an AGB star and a star-forming region is very low \citep{KastnerMyers1994}. For these reasons, the supernovae and stellar winds of massive stars, which yield SLRs much more quickly after star formation, are thought to be the most likely origin of ${}^{26}\mbox{Al}$ and ${}^{60}\mbox{Fe}$. Proposed mechanisms by which massive stars could enrich the infant Solar System fall into three broad scenarios: (1) supernova triggered collapse of pre-solar dense cloud core, (2) direct pollution of an already-formed proto-solar disc by supernova ejecta and (3) sequential star formation events in a molecular cloud.

The first scenario, supernova triggered collapse of pre-solar dense cloud core, was proposed by \citet{CameronTruran1977} just after the first discovery of ${}^{26}\mbox{Al}$ in Allende CAIs by \citet{LeeEtAl1976}. In this scenario, a nearby Type II supernova injects SLRs and triggers the collapse of the early Solar nebula. Many authors have simulated this scenario \citep{Boss1995, FosterBoss1996, BossEtAl2010, GritschnederEtAl2012, LiFrankBlackman2014, BossKeiser2014, Boss2017} and shown that it is in principle possible. A single supernova shock that encounters an isolated marginally stable prestellar core can compress it and trigger gravitational collapse while at the same time generating Rayleigh-Taylor instabilities at the surface that mix SLRs into the collapsing gas. However, these simulations have also demonstrated that this scenario requires severe fine-tuning. If the shock is too fast then it shreds and disperses the core rather than triggering collapse, and if it is too slow then mixing of SLRs does not occur fast enough to enrich the gas before collapse. Only a very narrow range of shock speeds are consistent with what we observe in the Solar System, and even then the SLR injection efficiency is low \citep{GritschnederEtAl2012, BossKeiser2014, Boss2017}. A possible solution to overcome the mixing barrier problem is the injection of SLRs via dust grains. However, only grains with radii larger than 30 $\rm \mu m$, which is much larger than the typical sizes of supernova grains (< 1 $\rm \mu m$), can penetrate the shock front and inject SLRs into the core \citep{BossKeiser2010}. Furthermore, analysis of Al and Fe dust grains in supernova ejecta constrains their sizes to be less than 0.01 $\rm \mu m$ \citep{BocchioEtAl2016}. \citet{DwarkadasEtAl2017} proposed a triggered star formation inside the shell of a Wolf-Rayet bubble, and found that the probability is from $0.01 - 0.16$.

The second scenario is a direct pollution: the Solar system's SLRs were injected directly into an already-formed protoplanetary disc by supernova ejecta within the same star-forming region \citep{Chevalier2000, HesterEtAl2004}. Hydrodynamical simulations of a protoplanetary disc have shown that the edge-on disc can survive the impact of a supernova blast wave, but that in this scenario only a tiny fraction of the supernova ejecta that strike the disc are captured and thus available to explain the SLRs we observe \citep{OuelletteDeschHester2007, ClosePittard2017}. \citet{OuelletteDeschHester2007} suggests that dust grains might be a more efficient mechanism for injecting SLRs into the disc, and simulations by \citet{OuelletteDeschHester2010} show that about 70 per cent of material in grains larger than 0.4 $\rm \mu m$ can be captured by a protoplanetary disc. However, extreme fine-tuning is still required to make this scenario work quantitatively. One can explain the observed SLR abundances only if SN ejecta are clumpy, the Solar nebula was struck by a clump that was unusually rich in ${}^{26}\mbox{Al}$ and ${}^{60}\mbox{Fe}$, and the bulk of these elements had condensed into large dust grains before reaching the Solar System. The probability that all these conditions are met is very low, $10^{-3} - 10^{-2}$. Moreover, the required dust size of 0.4 $\rm \mu m$ is still a factor of 40 larger than the value of 0.01 $\rm \mu m$ obtained by detailed study of dust grain properties by \citet{BocchioEtAl2016}.


The third scenario is sequential star formation events and self-enrichment in a giant molecular cloud (GMC) \citep{GounelleEtAl2009, GaidosEtAl2009, GounelleMeynet2012, Young2014, Young2016}. \citet{GounelleMeynet2012} proposed a detailed picture of this scenario; in a first star formation event, supernovae from massive stars inject ${}^{60}\mbox{Fe}$ to the GMC, and the shock waves trigger a second star formation event. This second star formation event also contains massive stars, and the stellar winds inject ${}^{26}\mbox{Al}$ and collect ISM gas to build a dense shell surrounding an H \textsc{ii} region. In the already enriched dense shell, a third star formation event occurs where the Solar system forms. \citet{VasileiadisEtAl2013} and \citet{KuffmeierEtAl2016} have modelled the evolution of a GMC by hydrodynamical simulations and shown that SN ejecta trapped within a GMC can enrich the GMC gas to abundance ratios of $^{26}\mbox{Al} / {}^{27}\mbox{Al}$ $\sim 10^{-6} - 10^{-4}$ and $^{60}\mbox{Fe} / {}^{56}\mbox{Fe}$ $\sim 10^{-7} - 10^{-5}$, comparable to or higher than any meteoritic estimates. However, this scenario requires that the bulk of the SLRs that are produced be captured within their parent GMCs. This is enforced by fiat in the simulations (by the use of periodic boundary conditions), but it is far from clear if this requirement can be met in reality. In the simulations the required enrichment levels are not reached for $\sim 15$ Myr, but observed young star clusters are always cleared of gas by ages of $\lesssim 5$ Myr \citep[e.g.,][]{hollyhead15a}. Moreover, the observed distribution of ${}^{26}\mbox{Al}$ has a scale height significantly larger than that of GMCs, which would seem hard to reconcile with the idea that most  ${}^{26}\mbox{Al}$ remains confined to the GMC where it was produced \citep{BouchetEtAl2015}. 

The literature contains a number of other proposals \citep[e.g.,][]{TatischeffDupratdeSereville2010, GoodsonEtAl2016}, but what they have in common with the three primary scenarios outlined above is that they require an unusual and improbable conjunction of circumstances (e.g., a randomly-passing WR star, SN-produced grains much larger than observations suggest) that would render the Solar System an unusual outlier in its abundances, or that they are not consistent with the observed distribution of $^{26}\mbox{Al}$ in the Galaxy.

Here we present an alternative scenario, motivated by two observations. First, ${}^{26}\mbox{Al}$ is observed to extend to a significant height above and below the Galactic disc, suggesting that regions contaminated by SLRs much be at least kpc-scale \citep{BouchetEtAl2015}. Second, there is no a priori reason why one should expect star formation to produce a SLR distribution with the same mean as the ISM as a whole, because star formation does not sample from the ISM at random. Instead, star formation and SLR production are both highly correlated in space and time \citep[e.g.,][]{EfremovElmegreen1998, GouliermisEtAl2010, GouliermisEtAl2015, GouliermisEtAl2017, GrashaEtAl2017a, GrashaEtAl2017b}; the properties of GMCs are also correlated on Galactic scales \citep[e.g.,][]{FujimotoEtAl2014, FujimotoEtAl2016, Colombo14a}. That both SLRs and star formation are correlated on kpc scales suggests that it is at these scales that we should search for a solution to the origin of SLRs in the early Solar System. 

In this paper, we will study the galactic-scale distributions of ${}^{26}\mbox{Al}$ and ${}^{60}\mbox{Fe}$ produced in stellar winds and supernovae, and propose a new contamination scenario: contamination due to Galactic-scale correlated star formation. In \autoref{Methods}, we present our numerical model of a Milky-Way like galaxy, along with our treatments of star formation and stellar feedback. In \autoref{Results}, we describe global evolution of the galactic disc and the abundance ratios of the stars that form in it. In \autoref{Discussion} we discuss the implications of our results, and based on them we propose a new scenario for SLR deposition. We summarise our findings in \autoref{Conclusions}.

\section{Methods}
\label{Methods}

We study the abundances of $^{60}\textrm{Fe}$ and $^{26}\textrm{Al}$ in newly-formed stars by performing a high-resolution chemo-hydrodynamical simulation of the interstellar medium (ISM) of a Milky-Way like galaxy. 
The simulation includes hydrodynamics, self-gravity, radiative cooling, photoelectric heating, stellar feedback in the form of photoionisation, stellar winds and supernovae to represent dynamical evolution of the turbulent multi-phase ISM, and a fixed axisymmetric logarithmic potential to represent the gravity of old stars and dark matter, which causes the galactic-scale shear motion of the ISM in a flat rotation curve. In the simulation, when self-gravity causes the gas to collapse past our ability to resolve, we insert "star particles" that represent stochastically-generated stellar populations drawn star-by-star from the initial mass function (IMF). Each massive star in these populations evolves individually until it produces a mass-dependent yield of $^{60}\mbox{Fe}$ and $^{26}\mbox{Al}$ at the end of its life. We subsequently track the transport and decay of these isotopes, and their incorporation into new stars. Further details on our numerical method are given in the following subsections.

We carry out all analysis and post-processing of the simulation outputs, and produce all simulations visualisations, using the \textsc{yt} software package \citep{TurkEtAl2011}.

\subsection{Chemo-hydrodynamical simulation}
\label{Chemo-hydrodynamical simulation}

Our simulations follow the evolution of a Milky-Way type galaxy using the adaptive mesh refinement code \textsc{enzo} \citep{BryanEtAl2014}. We use a piecewise parabolic mesh hydrodynamics solver to follow the motion of the gas. Since the $\sim 200\ \mathrm{km\ s^{-1}}$ circular velocity of the galaxy necessitates strongly supersonic flows in the galactic disc, we make use of the dual energy formalism implemented in the \textsc{enzo} code, in order to avoid spurious temperature fluctuations due to floating point round-off error when the kinetic energy is much larger than the internal energy. We treat isotopes as passive scalars that are transported with the gas, and that decay with half-lives of 2.62 Myr for $^{60}\mbox{Fe}$ and 0.72 Myr for $^{26}\mbox{Al}$ \citep{RugelEtAl2009, NorrisEtAl1983}. 

The gas cools radiatively to 10 K using a one-dimensional cooling curve created from the \textsc{cloudy} package's cooling table for metals and \textsc{enzo}'s non-equilibrium cooling rates for atomic species of hydrogen and helium \citep{AbelEtAl1997, FerlandEtAl1998}. This is implemented as tabulated cooling rates as a function of density and temperature \citep{JinEtAl2017}. In addition to radiative cooling, the gas can also be heated via diffuse photoelectric heating in which electrons are ejected from dust grains via FUV photons. This is implemented as a constant heating rate of $8.5 \times 10^{-26}\ \mathrm{erg\ s^{-1}}$ per hydrogen atom uniformly throughout the simulation box. This rate is chosen to match the expected heating rate assuming a UV background consistent with the Solar neighbourhood value \citep{Draine2011}. Self-gravity of the gas is also implemented.

We do not include dust grain physics because the typical drift velocity of the small dust ($\sim 0.1 \mu \rm m$) relative to gas at sub-parsec scale in the galactic disc is only $7.5 \times 10^{-4}\ \rm km/s$, much smaller than the typical turbulent velocity of the ISM ($\sim 10\ \rm km/s$) \citep{WibkingThompsonKrumholz2018}. Furthermore, analysis of Al and Fe dust grains in supernova ejecta constrains their sizes to be less than 0.01 $\rm \mu m$ \citep{BocchioEtAl2016}. Therefore, the dust grains and gas are very well coupled at the spatial scale we resolve in this simulation.

\subsection{Galaxy model}
\label{Galaxy model}

The galaxy is modelled in a three-dimensional simulation box of $(128\ \rm kpc)^3$ with isolated gravitational boundary conditions and periodic fluid boundaries. The root grid is $128^3$ with an additional 7 levels of refinement, producing a minimum cell size of 7.8125 pc. We refine a cell if the Jeans length, $\lambda_{\rm J} = c_{\rm s} \sqrt{\pi/(G\rho)}$, drops below 8 cell widths, comfortably satisfying the \citet{TrueloveEtAl1998} criterion. In addition, to ensure that we resolve stellar feedback, we require that any computational zone containing a star particle be refined to the maximum level. To keep the Jeans length resolved after collapse has reached the maximum refinement level, we employ a pressure floor such that the Jeans length is resolved by at least 4 cells on the maximum refinement level. In addition to the static root grid, we impose 5 additional levels of statically refined regions enclosing the whole galactic disc of 14 kpc radius and 2 kpc height. This guarantees that the circular motion of the gas in the galactic disc is well resolved, with a maximum cell size of 31.25 pc. 

We use initial conditions identical to those of \citet{TaskerTan2009}. These are tuned to the Milky-Way in its present state, but the Galaxy's bulk properties were not substantially different when Solar system formed 4.567 Gyr ago ($z \sim 0.4$). The simulated galaxy is set up as an isolated disc of gas orbiting in a static background potential which represents both dark matter and a stellar disc component. The form of the background potential is
\begin{equation}
\Phi (r, z) = \frac{1}{2} {v_{c, 0}^2} \ln \left[\frac{1}{{r_c^2}} \left( {r_c^2} + r^2 + \frac{z^2}{{q_{\phi}^2}} \right)\right],
\end{equation}
where $v_{c, 0}$ is the constant circular velocity at large radii, here set equal to 200 $\rm km\ s^{-1}$, $r$ and $z$ are the radial and vertical coordinates, the core radius is $r_c = 0.5\ \rm kpc$, and the axial ratio of the potential is $q_{\phi} = 0.7$. This corresponding circular velocity is
\begin{equation}
v_c = \frac{v_{c, 0} r}{\sqrt{{r_c^2} + r^2}}.
\end{equation}
The initial gas density distribution is
\begin{equation}
\rho (r, z) = \frac{\kappa \sqrt{{c_s^2} + {\sigma_{\rm 1D}^2}}}{2 \pi G Q z_h} {\rm sech}^2 \left( \frac{z}{z_h} \right),
\end{equation}
where $\kappa$ is the epicyclic frequency, $c_s$ is the sound speed, here set equal to $6\ \rm km\ s^{-1}$, $\sigma_{\rm 1D}$ is the one-dimensional velocity dispersion of the gas motions in the plane of the disc after the subtraction of the circular velocity, $Q$ is the Toomre stability parameter, and $z_h$ is the vertical scale height, which is assumed to vary with galactocentric radius following the observed radially-dependent H \textsc{i} scale height for the Milky-Way. Our disc is initialized with $\sigma_{1\rm D} = 0$. 

The initial disc profile is divided radially into three parts. In our main region, between radii of $r = 2-13\ \rm kpc$, $\rho$ is set so that $Q = 1$. The other regions of the galaxy, from 0 to 2 kpc and from 13 to 14 kpc, are initialised with $Q = 20$. Beyond 14 kpc, the disc is surrounded by a static, very low density medium. We set the initial abundances of $^{60}\mbox{Fe}$ and $^{26}\mbox{Al}$ to $10^{-12}$, though this choice has no practical effect since the initial abundances decay rapidly. In total, the initial gas mass is $8.6 \times 10^9\ M_{\odot}$, and the initial $^{60}\mbox{Fe}$ and $^{26}\mbox{Al}$ mass are set to $8.6 \times 10^{-3}\ M_{\odot}$.

Note that we do not include explicit spiral perturbations in our gravitational potential, but that flocculent spiral structure nonetheless forms spontaneously in our simulation as a result of gas self-gravity (see \autoref{Evolution of the Disc}). Similarly, we do not have a live model of the stellar bulge, but we implicitly include its effects on the gas via our potential, which has a bulge-like flattening at small radii. However, our simulation does not include the effects of a galactic bar, nor does it include the effects of cosmological inflow or tidal interactions with satellite galaxies. The influence of these effects should be addressed in a future work.

\subsection{Star formation}
\label{Star formation}

Implementations of star formation in galaxy-scale scale simulations such as ours are generally parameterised by two choices: a threshold density at which star formation begins, and an efficiency of star formation in cells above that threshold. In isolated galaxy simulations such as the one we perform, numerical experiments \citep[e.g.,][]{HopkinsEtAl2013} have shown that observed galaxies are best reproduced in simulations where the star formation threshold is set based on criteria of gravitational boundedness, i.e., star formation should occur only in fluid elements that are gravitationally bound or nearly so at the highest available numerical resolution. In a grid simulation such as ours, the criterion of boundedness is most conveniently expressed in terms of the ratio of the local Jeans length $\lambda_J$ to the local cell size $\Delta x$. We set our star formation threshold such that gas is star-forming if $\lambda_J / \Delta x < 4$ for $\Delta x$ at the maximum allowed refinement level \citep{TrueloveEtAl1997}; note that this choice guarantees that star formation occurs only in cells that have been refined to the highest allowed level. Rather than calculating the sound speed on the fly, it is more convenient to note that, at the densities at which we will be applying this condition, the gas is always very close to the thermal equilibrium defined by equality between photoelectric heating and radiative cooling (\autoref{Chemo-hydrodynamical simulation}). Consequently, we can reduce the condition for gas to be star-forming to a simple resolution-dependent density threshold by setting the sound speed based on the equilibrium temperature as a function of density. Doing so and plugging in the various resolutions we will use in this paper (see \autoref{Results}) yields number density thresholds for star formation of 12 $\rm cm^{-3}$ for a resolution $\Delta x = 31$ pc, 25.4 $\rm cm^{-3}$ for $\Delta x = 15$ pc and 57.5 $\rm cm^{-3}$ for $\Delta x = 8$ pc.

The second parameter in our star formation recipe characterises the star formation rate in gas that exceeds the threshold. We express the star formation rate density in cells that exceed the threshold as
\begin{equation}
\frac{d\rho_*}{dt} = \epsilon_{\rm ff} \frac{\rho}{t_{\rm ff}}.
\end{equation}
Here $\rho$ is the gas density of the cell, $t_{\rm ff} = \sqrt{3\pi / 32 G \rho}$ is the local dynamical time, and $
\epsilon_{\rm ff}$ is our rate parameter. Fortunately the value of $\epsilon_{\rm ff}$ is very well constrained by both observations and numerical experiments. For observations, one can measure $\epsilon_{\rm ff}$ directly by a variety of methods, and the consensus result from most techniques is that $\epsilon_{\rm ff} \approx 0.01$, with relatively little dispersion \citep[e.g.,][though see \citealt{Lee16a} for a contrasting view]{Krumholz07a, Krumholz12a, Evans14a, Heyer16a, Vutisalchavakul16a, Leroy17a, Onus18a}. From the standpoint of numerical experiments, a number of authors have shown that only simulations that fix $\epsilon_{\rm ff} \approx 0.01$ yield ISM density distributions consistent with observational constraints \citep[e.g.,][]{Hopkins13c, Semenov18a}. Given these constraints, we adopt $\epsilon_{\rm ff} = 0.01$ for this work.

To avoid creating an extremely large number of star particles whose mass is insufficient to have a well sampled stellar population, we impose a minimum star particle mass, $m_{\rm sf}$, and form star particles stochastically rather than spawn particles in every cell at each timestep. In this scheme, a cell forms a star particle of mass $m_{\rm sf} = 300\ M_{\odot}$ with probability
\begin{equation}
P = \left(\epsilon_{\rm ff} \frac{\rho}{t_{\rm ff}} \Delta x^3 \Delta t \right) / m_{\rm sf}, 
\end{equation}
where $\Delta x$ is the cell width, and $\Delta t$ is the simulation timestep. In practice, all star particles in our simulation are created via this stochastic method with masses equal to $m_{\rm sf}$. Note that the choice of the star particle of mass $300\ M_{\odot}$ does not affect the total star formation rate in the simulated galaxy as shown in Figure 1 in \citet{GoldbaumKrumholzForbes2015} , and we show \aref{Resolution} that our star particles are small enough that we resolve the characteristic size scale on which star formation is clustered extremely well, so that our choice of star particle mass does not affect the clustering of star formation either. Star particles are allowed to form in the main region of the disc between $2 < r < 14$ kpc.

\subsection{Stellar feedback}
\label{Stellar feedback}

Here we describe a subgrid model for star formation feedback that includes the effects of ionising radiation from young stars, the momentum and energy released by individual SN explosions, and gas and isotope injections from stellar winds and SNe. The inclusion of multiple forms of feedback is critical for producing results that agree with observations in high-resolution simulations such as ours \citep[e.g.,][]{Hopkins11a, AgertzEtAl2013, Stinson13a, Renaud13a}. In particular, simulations with enough resolution to capture the $\approx 5$ Myr delay between the onset of star formation and the first supernova explosions require non-supernova feedback in order to avoid overproducing stars (compared to what is observed) before supernovae have time to disperse star-forming gas. We pause here to note that this means that implementations of feedback are inevitably tuned to the resolution of the simulations being carried out, with simulations that go to higher resolution requiring the inclusion of more physical processes to replace the artificial softening of gravity that occurs at lower resolution. The feedback implementation we use here is tuned to the $\sim 10$ pc resolution we achieve, and is very similar to that of other authors who run simulations at similar resolution.

All star particles form with a uniform initial mass of 300 $M_{\odot}$. Within each of these particles we expect there to be a few stars massive enough to produce SN explosions. We model this using the \textsc{slug} stellar population synthesis code \citep{da-Silva12a, KrumholzEtAl2015}. This stellar population synthesis method is used dynamically in our simulation; each star particle spawns an individual \textsc{slug} simulation that stochastically draws individual stars from the initial mass function, tracks their mass- and age-dependent ionising luminosities, determines when individual stars explode as SNe, and calculates the resulting injection of $^{60}\mbox{Fe}$ and $^{26}\mbox{Al}$. In the \textsc{slug} calculations we use a Chabrier initial mass function \citep{Chabrier2005} with \textsc{slug}'s Poisson sampling option, Padova stellar evolution tracks with Solar metallicity \citep{GirardiEtAl2000}, \textsc{starburst99} stellar atmospheres \citep{LeithererEtAl1999}, and Solar metallicity yields from \citet{SukhboldEtAl2016}.

We include stellar feedback from photoionisation and SNe, following \citet{GoldbaumKrumholzForbes2016}, though our numerical implementation is very similar to that used by a number of previous authors \citep[e.g.,][]{Renaud13a}. For the former, we use the total ionising luminosity $S$ from each star particle calculated by \textsc{slug} to estimate the Str\"{o}mgren volume $V_s = S/\alpha_{\rm B} n^2$, and compare with the cell volume, $V_c$. Here $\alpha_{\rm B} = 2.6\times 10^{-13}$ cm$^3$ s$^{-1}$ is the case B recombination rate coefficient, $n = \rho/\mu m_{\rm H}$ is the number density, and $\mu = 1.27$ and $m_{\rm H} = 1.67 \times 10^{-24}$ g are the mean particle mass and the mass of an H nucleus, respectively. If $V_s < V_c$, the cell is heated to $10^4 (V_s/V_c)$ K. If $V_s > V_c$, the cell is heated to a temperature of $10^4$ K, and then we calculate the luminosity $S_{\rm esc} = S - \alpha_{\rm B} n^2 V_c$ that escapes the cell. We distribute this luminosity evenly over the neighbouring 26 cells, and repeat the procedure.

For SN feedback, a critical challenge in high resolution simulations such as ours is that the Sedov-Taylor radius for supernova remnants may or may not be resolved, depending on the ambient density in which the supernova explodes. In this regime several authors have carried out numerical experiments showing that the feedback recipes that best reproduce the results of high-resolution simulations are those that switch smoothly injecting pure radial momentum in cases where the Sedov-Taylor radius is unresolved to adding pure thermal energy in cases where it is resolved \citep[e.g.,][]{Kimm15a, Hopkins18c}. Our scheme, which is identical to that used in \citet{GoldbaumKrumholzForbes2016}, is motivated by this consideration. We identify particles that will produce SNe in any given time step. For each SN that occurs, we add a total momentum of $3 \times 10^5\ M_{\odot}\ \rm km\ s^{-1}$, directed radially outward in the 26 neighbouring cells. This momentum budget is consistent with the expected deposition from single supernovae \citep{GentryEtAl2017}. The total net increase in kinetic energy in the cells surrounding the SN host cell are then deducted from the available budget of $10^{51}$ erg and the balance of the energy is then deposited in the SN host cell as thermal energy. This scheme meets the requirement of smoothly switching from momentum to energy injection depending on the ambient density: if the explosion occurs in an already-evacuated region such that the gas density is low, the kinetic energy added in the process of depositing the radially outward momentum will be $\ll 10^{51}$ erg, and the bulk of the supernova energy will be injected as pure thermal energy. In a dense region, on the other hand, little thermal energy will remain, and only the radial momentum deposited will matter. In the higher resolution phases of the simulation ($\Delta x =$ 15 pc, 8 pc), we increase the momentum budget to $5 \times 10^5\ M_{\odot}\ \rm km\ s^{-1}$ in order to maintain approximately the same total star formation rate; given that the actual momentum budget is uncertain by a factor of $\approx 10$ due to the effects of clustering \citep{GentryEtAl2017}, this value is still well within the physically plausible range.

We include gas mass injection from stellar winds and SNe to each star particle's host cell each time step. The mass loss rate of each star particles is calculated from the \textsc{slug} stellar population synthesis. Note that we do not include energy injection from stellar winds; these will be included in future work. However, even though the simulation does not include the effect, the total star formation rate in the simulated galaxy is consistent with observations.

We include isotope injection from stellar winds and SNe, which is calculated from the mass-dependent yield tables of \citet{SukhboldEtAl2016}. The explosion model for massive stars is one-dimensional, of a single metallicity (solar) and does not include any effects of stellar rotation. The chemical yields are deposited to the host cell. As discussed in \citet{SukhboldEtAl2016}, their nucleosynthesis model overpredicts\footnote{\citet{SukhboldEtAl2016} compared their ejected mass ratio of $^{60}\mbox{Fe}/{}^{26}\mbox{Al}$ (= 0.9) with the observed steady-state mass ratio of 0.34 \citep{WangEtAl2007}, and stated that their yield should be corrected by a factor of three. However, a steady-state mass ratio should be used, not the ejected mass ratio, to compare with the observed mass ratio. The steady-state mass ratio can be obtained by multiplying the ratio of half-lives, as $0.9 \times (2.62\ {\rm Myr}/0.72\ {\rm Myr}) = 3.3$. This steady-state mass is ten times larger than the observed steady-state mass ratio. That is why we modify their tables by reducing the $^{60}\mbox{Fe}$ yield by a factor of five and doubling the $^{26}\mbox{Al}$ yield.} the $^{60}\mbox{Fe}$ to $^{26}\mbox{Al}$ compared to that determined from $\gamma$-ray line observations \citep{WangEtAl2007}. 
They note that the discrepancy might have to do with errors in poorly-known nuclear reaction rates, especially for $^{26}\mbox{Al}(n, p)^{26}\mbox{Mg}$, $^{26}\mbox{Al}(n, \alpha)^{23}\mbox{Na}$, $^{59, 60}\mbox{Fe}(n, \gamma)^{60, 61}\mbox{Fe}$, or with uncertainties in stellar mixing parameters such as the strength of convective overshoot. Rotational mixing is another possible effect that is not considered in their chemical yields \citep{ChieffiLimongi2013, LimongiChieffi2018}.
To ensure that our $^{60}\mbox{Fe}/{}^{26}\mbox{Al}$ ratio is consistent with observations, we modify their tables slightly by reducing the $^{60}\mbox{Fe}$ yield by a factor of five and doubling the $^{26}\mbox{Al}$ yield. This brings our Galaxy-averaged ratios of $^{60}\mbox{Fe}/{}^{26}\mbox{Al}$, $^{60}\mbox{Fe}/\mbox{SFR}$, and $^{26}\mbox{Al}/\mbox{SFR}$ into good agreement with observations. Although uncertainties in the chemical yields might affects our results, we expect the effect to be at most a factor of ten, not orders of magnitude, since this is the current level of discrepancy between the numerical results and the observations. It would be worthwhile repeating our simulations in the future with other models of chemical yields \citep{EkstromEtAl2012, LimongiChieffi2006, LimongiChieffi2018, ChieffiLimongi2013, NomotoEtAl2006, NomotoKobayashiTominaga2013, PignatariEtAl2016}.

\section{Simulation Results}
\label{Results}

\subsection{Evolution of the Disc}
\label{Evolution of the Disc}

\begin{figure}
\includegraphics[width=\columnwidth]{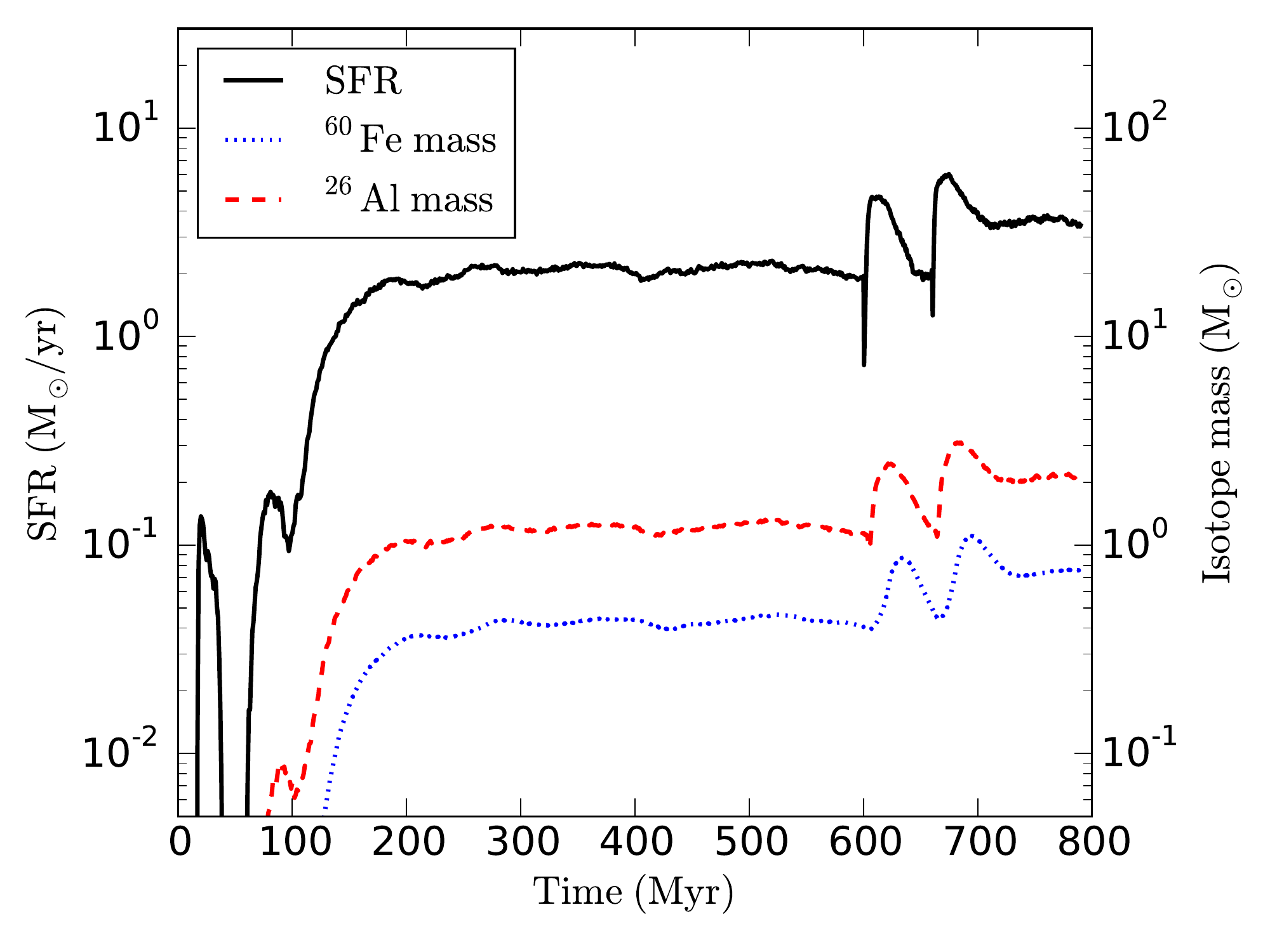}
\caption{The time evolution of SFR and isotope mass. The black solid line shows the total SFR in the galactic disc. The blue dotted and red dashed lines show the total mass of $\rm ^{60}Fe$ and $\rm ^{26}Al$ respectively. The sharp features at 600 and 660 Myr are transients caused when we increase the resolution.}
\label{fig:time_evolution}
\end{figure}

\begin{figure*}
\includegraphics[width=\hsize]{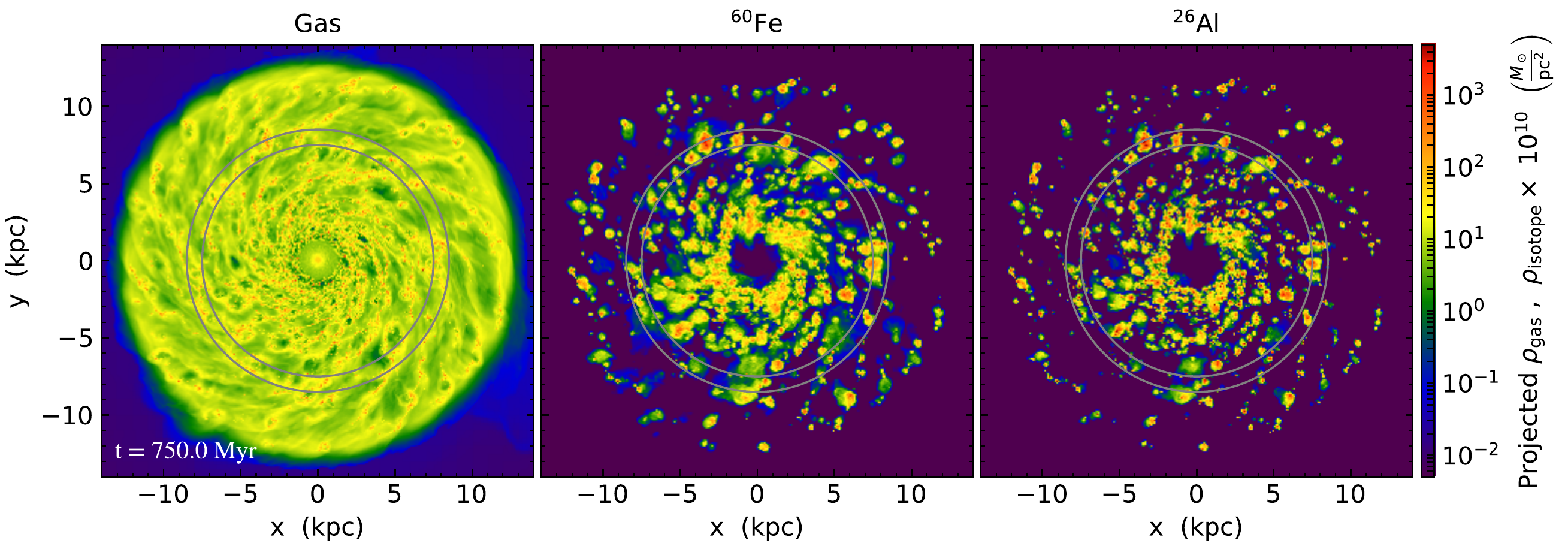}
\caption{The morphology of the galactic disc. Panels show the gas (left), $\rm ^{60}Fe$ (middle) and $\rm ^{26}Al$ (right) surface densities of the face-on disc at $t = $ 750 Myr. Each image is 28 kpc across. The galactic disc rotates anticlockwise. The two circles indicate Galactocentric radii of 7.5 kpc and 8.5 kpc, roughly bounding the Solar annulus.}
\label{fig:density_projection}
\end{figure*}

\begin{figure*}
\includegraphics[width=\hsize]{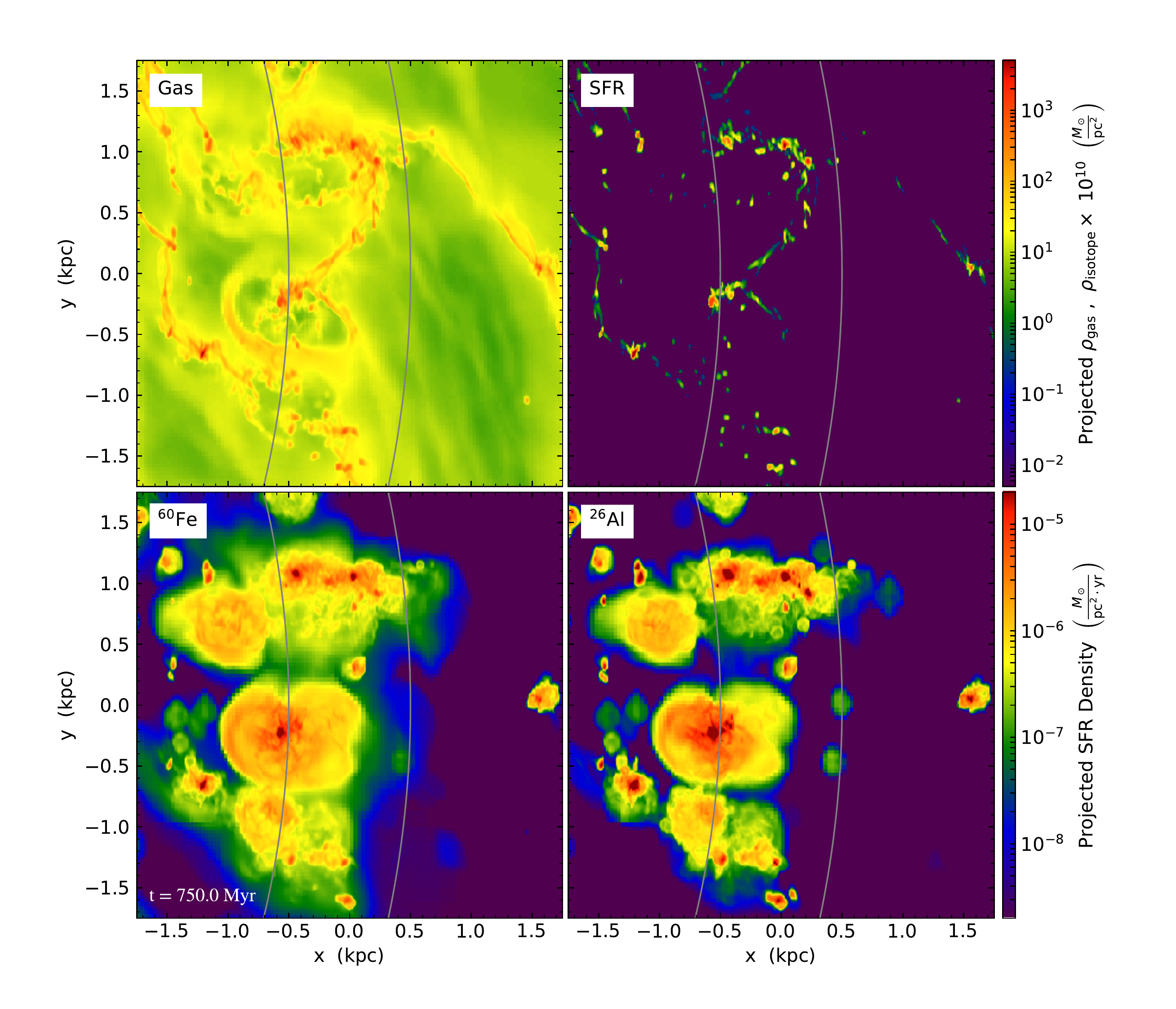}
\caption{Same as \autoref{fig:density_projection}, but zoomed in on a spot near the Solar Circle. Panels show the gas (top-left), star formation rate (top-right), $\rm ^{60}Fe$ (bottom-left) and $\rm ^{26}Al$ (bottom-right) surface densities at $t = $ 750 Myr. The two arcs show Galactocentric radii of 7.5 and 8.5 kpc, bounding the Solar annulus.}
\label{fig:density_projection_zoom}
\end{figure*}

To determine the equilibrium distributions of isotopes in newly-formed stars, we use a relaxation strategy to allow the simulated galaxy to settle into statistical equilibrium at high resolution. We first run the simulation at a resolution of 31 pc for 600 Myr, corresponding to two rotation periods at 10 kpc from the galactic centre. This time is sufficient to allow the disc to settle into statistical steady state, as we illustrate in \autoref{fig:time_evolution}, which shows the time evolution of the total star formation rate (SFR) and total $^{60}\textrm{Fe}$ and $^{26}\textrm{Al}$ masses within the Galaxy. We then increase the resolution from 31 pc to 15 pc and allow the disc to settle back to steady state at the new resolution, which takes until 660 Myr. At that point we increase the resolution again, to 8 pc. These refinement steps are visible in \autoref{fig:time_evolution} as sudden dips in the SFR, which occur because it takes some time after we increase the resolution for gas to collapse past the new, higher star formation threshold, followed by sudden bursts as a large mass of gas simultaneously reaches the threshold. However, feedback then pushes the system back into equilibrium. In the equilibrium state the SFR is $1-3$ $M_\odot$ yr$^{-1}$, consistent with the observed Milky-Way star formation rate \citep{ChomiukPovich2011}. Similarly, the total SLR masses in the equilibrium state are 0.7 $\rm M_\odot$ for $\rm ^{60}Fe$ and 2.1 $\rm M_\odot$ for $\rm ^{26}Al$, respectively, consistent with masses determined from $\gamma$-ray observations \citep{Diehl2017, WangEtAl2007}.  Note that, as we change the resolution, the steady-state SFR and SLR abundances vary at the factor of $\approx 2$ level. This is not surprising, because our stellar feedback model operates on a stencil of $3^3$ cells around each star particles, and thus volume over which we inject feedback varies as does the resolution. However, we note that the variations in equilibrium SFR and SLR mass with resolution are well within both the observational uncertainties on these quantities.

\autoref{fig:density_projection} shows the global distributions of gas and isotopes in the galactic disc at $t = $ 750 Myr, when the maximum resolution is 8 pc and the galactic disc is in a quasi-equilibrium state. \autoref{fig:density_projection_zoom} shows the same data, zoomed in on a $3.5$ kpc-region centred on the Solar Circle\footnote{Simulation movies are available at \url{https://sites.google.com/site/yusuke777fujimoto/data}}. The Figures show that the disc is fully fragmented, and has produced GMCs and star-forming regions.
The distributions of $\rm ^{60}Fe$ and $\rm ^{26}Al$ are strongly-correlated with the star-forming regions, which correspond to the highest-density regions (reddish colours) visible in the gas plot.
This is as expected, since these isotopes are produced by massive stars, which, due to their short lives, do not have time to wander far from their birth sites.

However, there are important morphological differences between the distributions of $\rm ^{60}Fe$, $\rm ^{26}Al$, and star formation. The $^{60}\mbox{Fe}$ distribution is the most extended, with the typical region of $^{60}\mbox{Fe}$ enrichment exceeding 1 kpc in size, compared to $\sim 100$ pc or less for the density peaks that represent star-forming regions. The $^{26}\mbox{Al}$ distribution is intermediate, with enriched regions typically hundreds of pc in scale. The larger extent of $^{60}\mbox{Fe}$ compared to $^{26}\mbox{Al}$ is due to its larger lifetime (2.62 Myr versus 0.72 Myr for $^{26}$Al) and its origin solely in fast-moving SN ejecta (as opposed to pre-SN winds, which contribute significantly to $^{26}$Al).

\begin{figure*}
\includegraphics[width=\hsize]{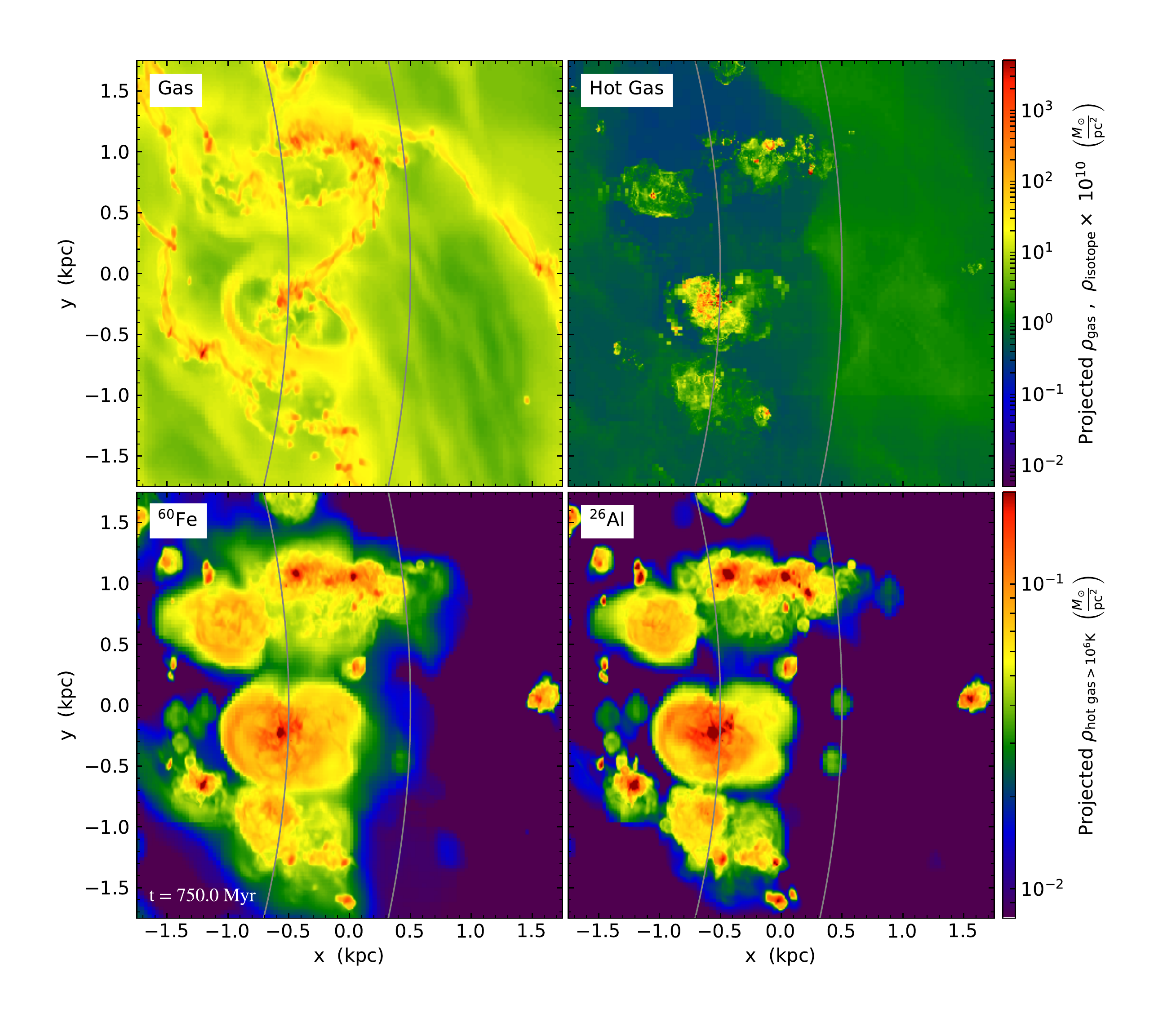}
\caption{Same as \autoref{fig:density_projection_zoom}, but showing hot gas ($> 10^6\ \mathrm{K}$) on the top-right panel.}
\label{fig:hot_gas_projection_zoom}
\end{figure*}

In addition to the comparison between SLRs and star formation, it is interesting to compare SLRs to the distribution of hot gas produced by supernovae (defined here as gas with temperature $T>10^6$ K), which we show in \autoref{fig:hot_gas_projection_zoom}. We see that, as expected, regions of $^{60}\mbox{Fe}$ and $^{26}\mbox{Al}$ enrichment correlate well with bubbles of hot gas. However, it is interesting to note that the outer edges of the $^{60}\mbox{Fe}$ or $^{26}\mbox{Al}$ bubbles seen in \autoref{fig:hot_gas_projection_zoom} extend significantly further than the bubbles of hot ISM. This could be a result either of cooling of the hot gas on timescales shorter than the decay of SLRs, or of rapid mixing of SLRs into cooler regions. Regardless, our finding that regions of SLR enrichment are generally larger in extent than regions of hot gas may be testable in the future as higher resolution observations of $\gamma$-ray emission from SLRs observed \textit{in situ} in the ISM become available.

\subsection{Abundance ratios in newborn stars}

\begin{figure}
\includegraphics[width=\columnwidth]{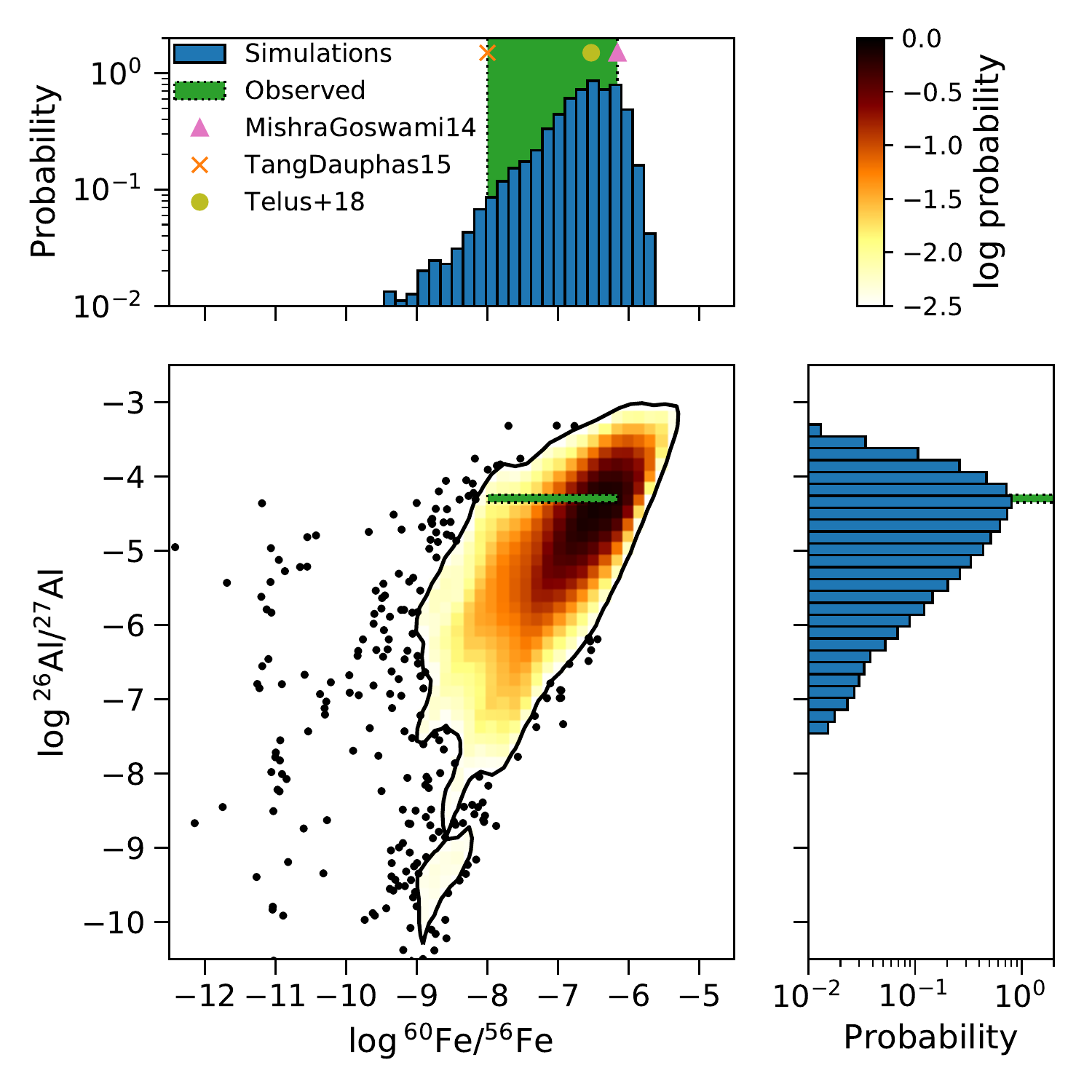}
\caption{
The abundance ratios of short-lived isotopes in newly-formed stars. The central panel shows the joint PDF of $\rm ^{60}Fe/^{56}Fe$ and $\rm ^{26}Al/^{27}Al$ from our simulations, with colours showing probability density and black points showing individual stars in sparse regions. The top and right panels show the PDFs of $\rm ^{60}Fe/^{56}Fe$ and $\rm ^{26}Al/^{27}Al$ individually, with simulations shown in blue. All simulation data are for stars formed from $740 - 750$ Myr, at Galactocentric radii from 7.5 - 8.5 kpc. Green bands show the uncertainty range of Solar System meteoritic abundances \citep{LeeEtAl1976, MishraGoswami2014, TangDauphas2015, TelusEtAl2018}; for $^{60}\mbox{Fe}$, due to the wide range of values reported in the literature, we also show three representative individual measurements as indicated in the legend.
}
\label{fig:abundance_ratios}
\end{figure}

To investigate abundance ratios of isotopes in newborn stars, whenever a star particle forms in our simulations, we record the abundances of $\rm ^{60}Fe$ and $\rm ^{26}Al$ in the gas from which it forms, since these should be inherited by the resulting stars. We do not add any additional decay, because our stochastic star formation prescription does not immediately convert gas to stars as soon as it crosses the density threshold, and instead accounts for the finite delay between gravitational instability and final collapse. 
\autoref{fig:abundance_ratios} shows the probability distribution functions (PDFs) for the abundance ratios $^{60}\mbox{Fe} / {}^{56}\mbox{Fe}$ and $^{26}\mbox{Al} / {}^{27}\mbox{Al}$; we derive the masses of the stable isotopes ${}^{56}\mbox{Fe}$ and ${}^{27}\mbox{Al}$ from the observed abundances of those species in the Sun \citep{AsplundEtAl2009}, and we measure the PDFs for star particles that form between 740 and 750 Myr in the simulation, at galactocentric radii from $7.5 - 8.5$ kpc (i.e., within $\approx 0.5$ kpc of the Solar Circle). However, the results do not strongly vary with galactocentric radius, as shown in \aref{Radial dependence}. We also show that the PDFs are converged with respect to spatial resolution at their high-abundance ends (though not on their low-abundance tails) in \aref{Resolution}.

In \autoref{fig:abundance_ratios} we also show meteoritic estimates for these abundance ratios \citep{LeeEtAl1976, MishraGoswami2014, TangDauphas2015, TelusEtAl2018}.
The PDF of $\rm ^{60}Fe$ peaks near $^{60}\mbox{Fe} / {}^{56}\mbox{Fe} \sim 3\times10^{-7}$, but is $\sim 2$ orders of magnitude wide, placing all the meteoritic estimates well within the ranges covered by the simulated PDF. The $^{26}\mbox{Al}$ abundance distribution is similarly broad, but the measured meteoritic value sits very close to its peak, as $^{26}\mbox{Al} / {}^{27}\mbox{Al} \sim 5\times10^{-5}$.
Clearly, the abundance ratios measured in meteorites are fairly typical of what one would expect for stars born near the Solar Circle, and thus the Sun is not atypical.

\section{Discussion}
\label{Discussion}

Our simulations suggest a mechanism by which the SLRs came to be in the primitive Solar System that is quite different than proposed in earlier work based on smaller-scale simulations or analytic models. We call this new contamination scenario ``inheritance from Galactic-scale correlated star formation". Our scenario differs substantially from the triggered collapse or direct injection scenarios in that both of these require unusual circumstances -- the core that forms the Sun is either at just the right distance from a supernova to be triggered into collapse but well-mixed, or the protoplanetary disc was hit by supernova ejecta and managed to capture them without being destroyed. In either case stars with SLR abundances like those of the Solar System should be rare outliers, while we find that the Sun's abundances are typical.

\begin{figure*}
\includegraphics[width=\hsize]{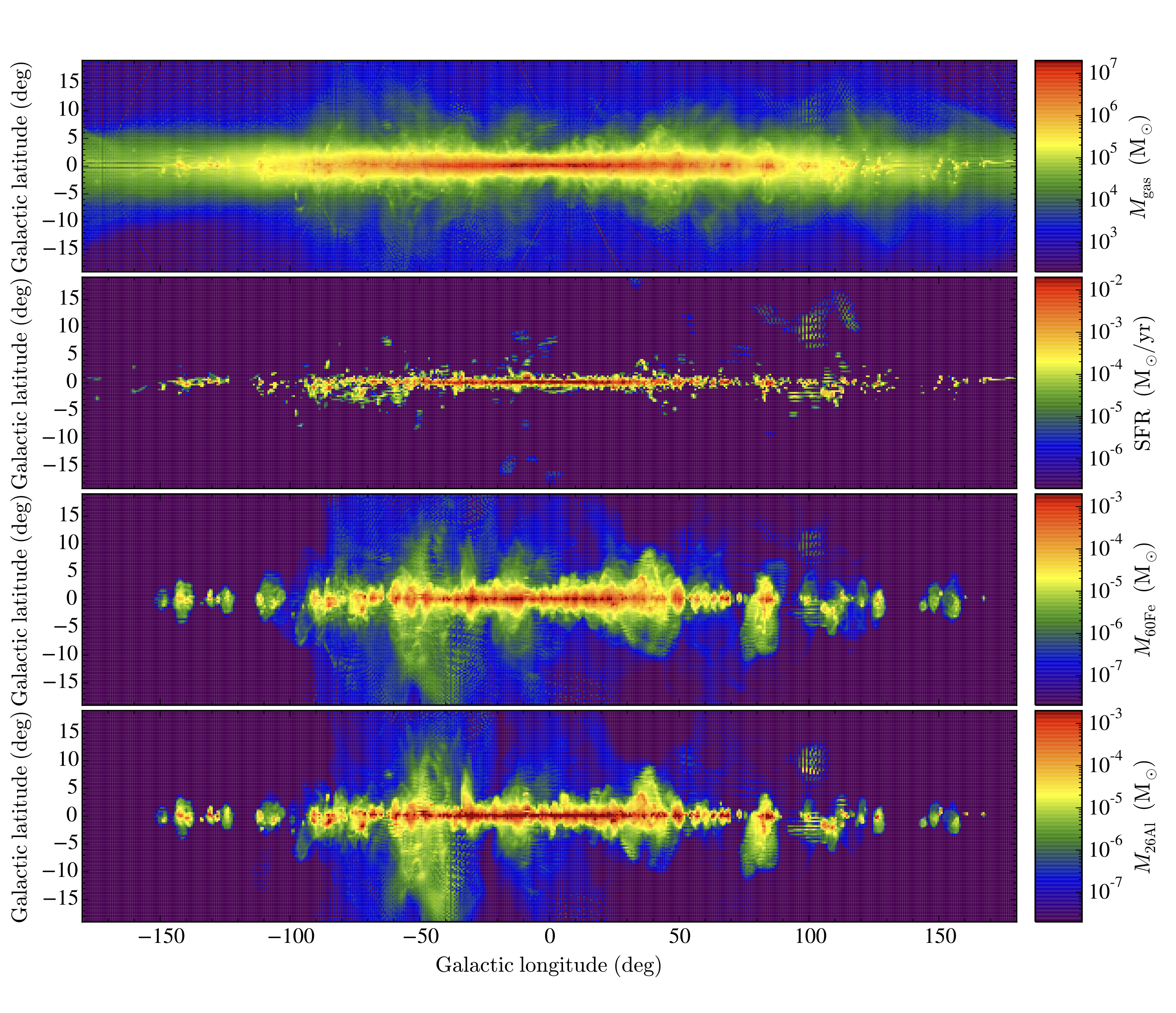}
\caption{Distributions of gas, star formation, and SLRs in Galactic coordinates, as viewed from the position of the Sun (i.e., a point 8 kpc from the Galactic centre). Panels show the gas, star formation rate, $\rm ^{60}Fe$ and $\rm ^{26}Al$ distributions (from top to bottom) in Galactic coordinates. Note that, although the absolute scales on the colour bars in each panel differ, all panels use the same dynamic range, and thus the distributions are directly comparable. The scalloping pattern that is visible at high latitudes and toward the outer galaxy is an artefact due to aliasing between the Cartesian grid and the angular coordinates in regions where the resolution is low.}
\label{fig:l_b_plot}
\end{figure*}

However, the scenario illustrated in our simulations is also very different from the GMC confinement hypothesis. To see why, one need only examine \autoref{fig:density_projection_zoom}. Observed GMCs, and those in our simulations, are at most $\sim 100$ pc in size, whereas in \autoref{fig:density_projection_zoom} we clearly see that regions of $^{60}\mbox{Fe}$ and $^{26}\mbox{Al}$ contamination are an order of magnitude larger. This difference between our simulations and the GMC confinement hypothesis is also visible in the distribution of $^{26}\mbox{Al}$ on the sky as seen from Earth. \autoref{fig:l_b_plot} shows all-sky maps of the gas, star formation rate, $^{60}\mbox{Fe}$, and $^{26}\mbox{Al}$ as viewed from a point 8 kpc from the Galactic Centre (i.e., at the location of the Sun). We should not regard \autoref{fig:l_b_plot} as an exact prediction of the $\gamma$-ray sky as seen from Earth, since we have not taken care to replicate the Sun's placement relative to spiral arms, nor have we tried to match the sky positions of local structures such as the Sco-Cen association that may have a large impact on what we observe from Earth. However, it is nonetheless interesting to examine the large-scale qualitative behaviour of the map shown in \autoref{fig:l_b_plot}, and its implications. If SLRs are confined by GMCs, then $\gamma$-rays from $^{26}$Al decay should have an angular thickness on the sky comparable to that of star-forming regions. \autoref{fig:l_b_plot} clearly shows that this is not the case in our simulations: $^{60}\mbox{Fe}$ and $^{26}\mbox{Al}$ extend to galactic latitude $b = 4^{\circ} - 5^{\circ}$, while star forming regions are confined to $b < 2^{\circ}$. The difference in scale heights we find is consistent with observations. The Galactic CO survey of \citet{DameEtAl2001} finds that most emission is confined to Galactic latitudes $b < 2^{\circ}$, while the $\gamma$-ray emission maps of $^{26}\mbox{Al}$ \citep{PluschkeEtAl2001, BouchetEtAl2015} show a thick disc with $b \approx 5^{\circ}$. Our simulation successfully reproduces the observed difference in $^{26}\mbox{Al}$ and CO angular distribution.

\begin{figure*}
\includegraphics[width=\hsize]{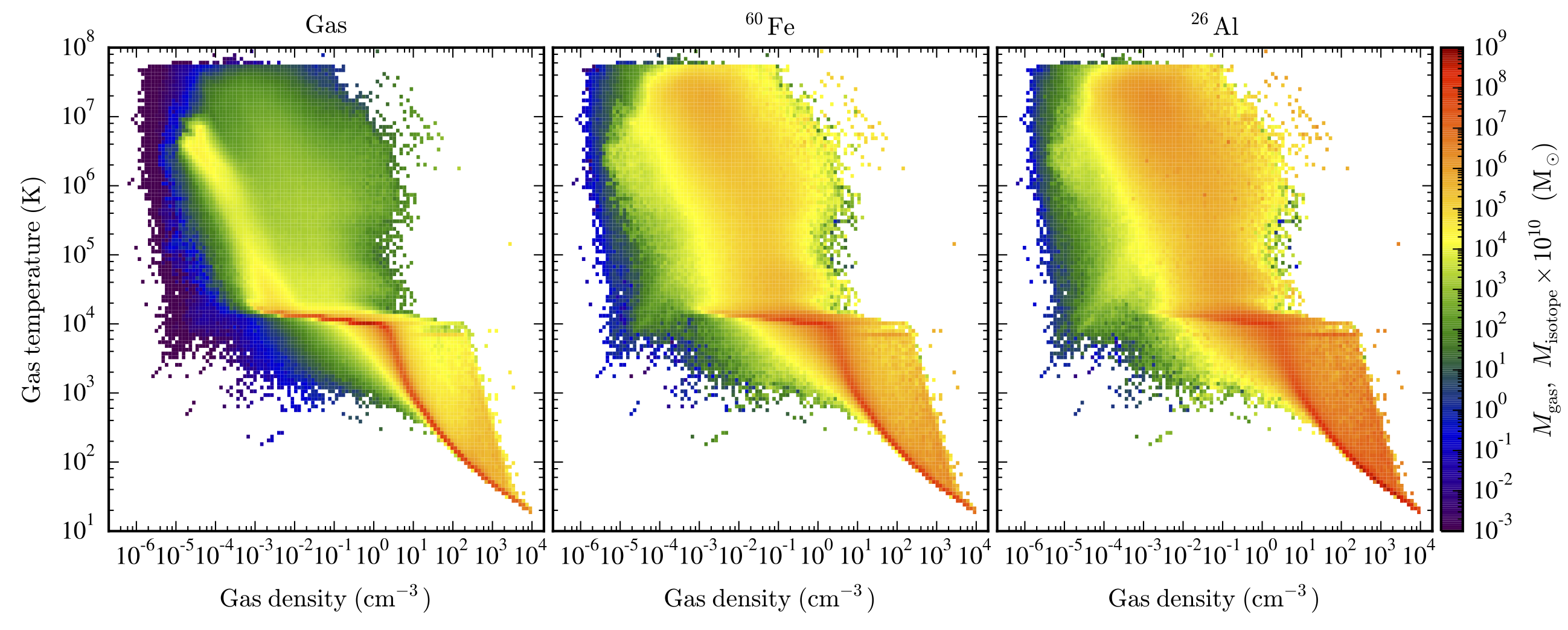}
\caption{Mass distributions with respect to gas temperature versus density. Left is the gas, middle is $\rm ^{60}Fe$ and right is $\rm ^{26}Al$ at $t = $ 750 Myr.}
\label{fig:phase_plots}
\end{figure*}

We can make this discussion more quantitative by examining the distribution of $^{60}\mbox{Fe}$ and $^{26}\mbox{Al}$ and their correlation with the gas and star formation properties of the galaxy. We first examine the distribution of the SLRs with respect to gas density and temperature, as illustrated in \autoref{fig:phase_plots}. We find that only 30\% of the $^{60}\mbox{Fe}$ and 56\% of the $^{26}\mbox{Al}$ by mass are found in GMCs (defined as gas with a density above 100 H cm$^{-3}$), compared to a total GMC mass fraction of 16\%; thus $^{60}\mbox{Fe}$ is overabundant in GMCs compared to the bulk of the ISM by less than a factor of 2, and $^{26}\mbox{Al}$ by less than a factor of 3.5. These modest enhancements are inconsistent with the hypothesis that SLRs abundances are high in the Solar System because SLRs are trapped within long-lived GMCs.

\begin{figure}
\includegraphics[width=\columnwidth]{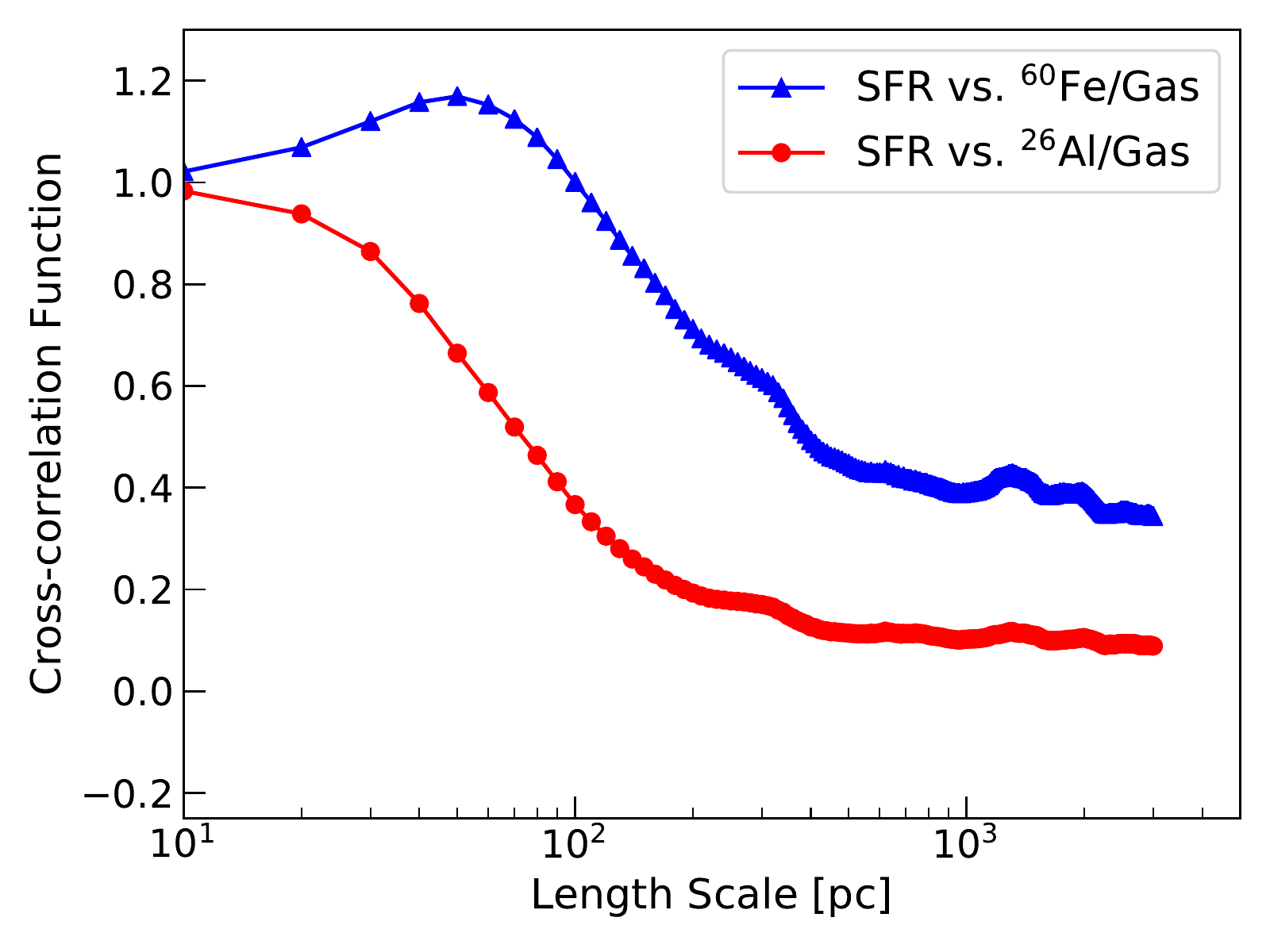}
\caption{Normalised spatial cross-correlations $(f * g) (r)$ between the SFR surface density and the surface densities of $^{60}\mbox{Fe}$ and $^{26}\mbox{Al}$ divided by the gas surface density.}
\label{fig:cross_correlation}
\end{figure}

We can also reach a similar conclusion by examining the spatial correlation of star formation with SLRs. For any two-dimensional fields $f(\vec{r})$ and $g(\vec{r})$ defined as a function of position $\vec{r}$ within the galactic disc, we can define the normalised spatial cross-correlation function $(f * g) (r)$ as
\begin{equation}
\label{eq:correlation_function}
(f * g) (r) = \frac{\left\langle \int f(\vec{r}') g(\vec{r}'-\vec{r})\, d\vec{r}' \right\rangle}{\int f(\vec{r}') g(\vec{r}')\, d\vec{r}'}
\end{equation}
where $r = |\vec{r}|$, and the angle brackets indicate an average over all possible angles of the displacement vector $\vec{r}$. In practice we can compute the correlation numerically using projected images such as those shown in \autoref{fig:density_projection} for two quantities $f$ and $g$. The denominator is simply the product of the two images, while we can obtain the integral in the numerator for a displacement vector $\vec{r}$ by shifting one of the images by $\vec{r}$, multiplying the shifted and unshifted images, and measuring product of the two images. We then compute the average over angle by averaging the numerator over shifts of the same magnitude $r = |\vec{r}|$. We show the spatial cross-correlation between star formation and element abundance ratios in \autoref{fig:cross_correlation}. As one can see from the figure, star formation is correlated with $^{60}\mbox{Fe}$ abundance on scales of 1 kpc and $^{26}\mbox{Al}$ abundance on scales of hundreds of pc, much larger than an individual GMC or star-forming complex. The difference of the correlation scales between the $^{60}\mbox{Fe}$ and $^{26}\mbox{Al}$ comes from the different lifetimes (2.62 Myr versus 0.72 Myr) and the fact that $^{60}\mbox{Fe}$ is added to the ISM only through fast-moving SN ejecta, while $^{26}\mbox{Al}$ has contributions from both supernovae and pre-SN stellar winds. This is consistent with the different morphological distributions of $^{60}\mbox{Fe}$ and $^{26}\mbox{Al}$ as  shown in \autoref{fig:density_projection_zoom}. The results do not strongly vary with galactocentric radius, as shown in \aref{Radial dependence}.

The overall picture that emerges from our simulations is that SLR abundances in newborn stars are large because star formation is highly-correlated in time and space \citep{EfremovElmegreen1998, GouliermisEtAl2010, GouliermisEtAl2015, GouliermisEtAl2017, GrashaEtAl2017a, GrashaEtAl2017b}. SN ejecta are not confined to individual molecular clouds, and instead deposit radioactive isotopes in the atomic gas over $\sim 1$ kpc from their parent molecular clouds. However, because star formation is correlated, and because molecular clouds are not closed boxes but instead continually accrete the atomic gas during their star forming lives \citep{FukuiKawamura2010, GoldbaumEtAl2011, Zamora-AvilesVazquez-SemadeniColin2012}, the pre-enriched atomic gas within $\sim 1$ kpc of a molecular cloud stands a far higher chance of being incorporated into a molecular cloud and thence into stars within a few Myr than does a random portion of the ISM at similar density and temperature. Conversely, the atomic gas in a galaxy that will be incorporated into a star a few Myr in the future does not represent an unbiased sampling of all the atomic gas in the galaxy. Instead, it is preferentially that atomic gas that is close to sites of current star formation, and thus is far more likely than average to have been contaminated with SLRs. It is the galactic scale correlation of star formation that is the key physical mechanism that produces high SLR abundances in the primitive Solar System and other young stars.

\section{Conclusions}
\label{Conclusions}

Short-lived radioisotopes (SLRs) such as $^{60}\mbox{Fe}$ and $^{26}\mbox{Al}$ are radioactive elements with half-lives less than 15 Myr that studies of meteorites have shown to be present at the time when the most primitive Solar System bodies condensed. The most likely origin site for the $^{60}\mbox{Fe}$ and $^{26}\mbox{Al}$ in meteorites is nucleosynthesis in massive stars, but the exact delivery mechanism by which these elements entered the Solar System's protoplanetary disc are still debated. 

To address this question, we have performed the first chemo-hydrodynamical simulation of the entire Milky-Way Galaxy (\autoref{fig:density_projection}), including stochastic star formation and stellar feedback in the form of H \textsc{ii} regions, supernovae, and element injection. Our simulations have enough resolution to capture individual supernovae, so that we can properly measure the full range of variation in SLR abundances that results from the stochastic nature of element production and transport. From our simulations we measure the expected distribution of $^{60}\mbox{Fe} / {}^{56}\mbox{Fe}$ and $^{26}\mbox{Al} / {}^{27}\mbox{Al}$ ratios for all stars in the Galaxy (\autoref{fig:abundance_ratios}). We find that the Solar abundance ratios inferred from meteorites are well within the normal range for Milky-Way stars; contrary to some models for the origins of SLRs, the Sun's SLR abundances are not atypical.

Our results lead us to propose a new enrichment scenario: SLR enrichment via Galactic-scale correlated star formation. We find that GMCs are at most 100 pc in size and their star forming regions are much smaller, while regions of $^{60}\mbox{Fe}$ and $^{26}\mbox{Al}$ contamination due to supernovae are an order of magnitude larger (\autoref{fig:density_projection_zoom}). The extremely broad distribution of $^{26}\mbox{Al}$ produced in our simulations is consistent with the observed distribution on the sky, which shows an angular scale height that is close to twice that of the molecular gas and star formation in the Milky-Way (\autoref{fig:l_b_plot}). The SLRs are not confined to the molecular clouds in which they are born (\autoref{fig:phase_plots}). However, SLRs are nonetheless abundant in newborn stars because star formation is correlated on galactic scales (\autoref{fig:cross_correlation}). Thus, although SLRs are not confined, they are in effect pre-enriching a halo of the atomic gas around existing GMCs that is very likely to be subsequently accreted or to form another GMC, so that new generations of stars preferentially form in patches of the Galaxy contaminated by previous generations of stellar winds and supernovae.

In future work, we will extend our simulations to include other SLRs such as $^{41}\mbox{Ca}$ and $^{53}\mbox{Mn}$, which also have been claimed to place severe constraints on the birth environment of the Solar system \citep{HussEtAl2009}.

\section*{Acknowledgements}

The authors would like to thank the referee, Roland Diehl, for his careful reading and helpful suggestions. Simulations were carried out on the Cray XC30 at the Center for Computational Astrophysics (CfCA) of the National Astronomical Observatory of Japan and the National Computational Infrastructure (NCI), which is supported by the Australian Government. Y.F. and M.R.K.~acknowledge support from the Australian Government through the Australian Research Council's \textit{Discovery Projects} funding scheme (project DP160100695). Computations described in this work were performed using the publicly available \textsc{enzo} code (\citealt{BryanEtAl2014}; \url{http://enzo-project.org}), which is the product of a collaborative effort of many independent scientists from numerous institutions around the world. Their commitment to open science has helped make this work possible. We acknowledge extensive use of the \textsc{yt} package (\citealt{TurkEtAl2011}; \url{http://yt-project.org}) in analysing these results and the authors would like to thank the \textsc{yt} development team for their generous help.




\bibliographystyle{mnras}
\bibliography{reference} 




\appendix

\section{Radial dependence}
\label{Radial dependence}

\begin{figure*}
\includegraphics[width=\columnwidth]{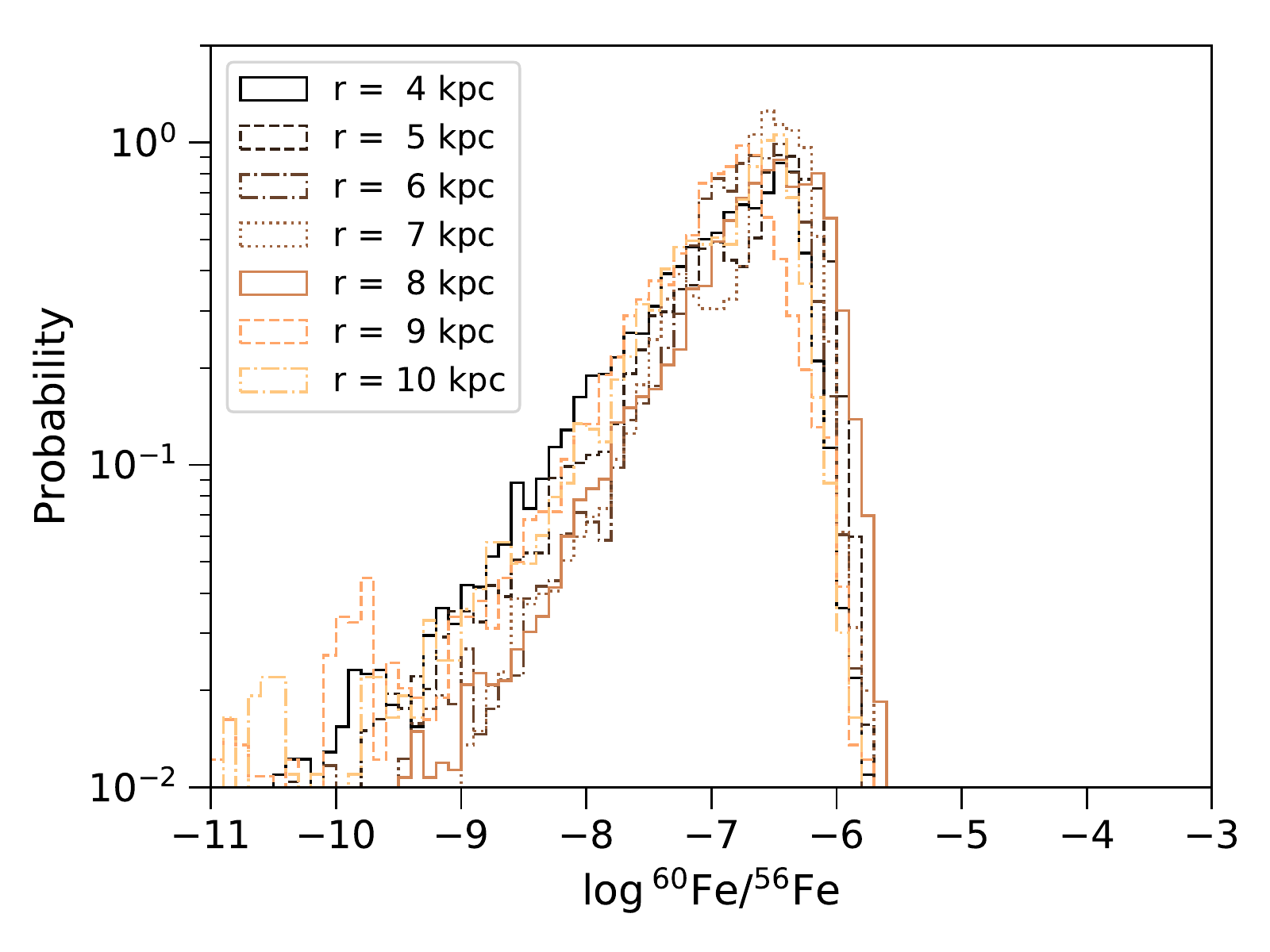}
\includegraphics[width=\columnwidth]{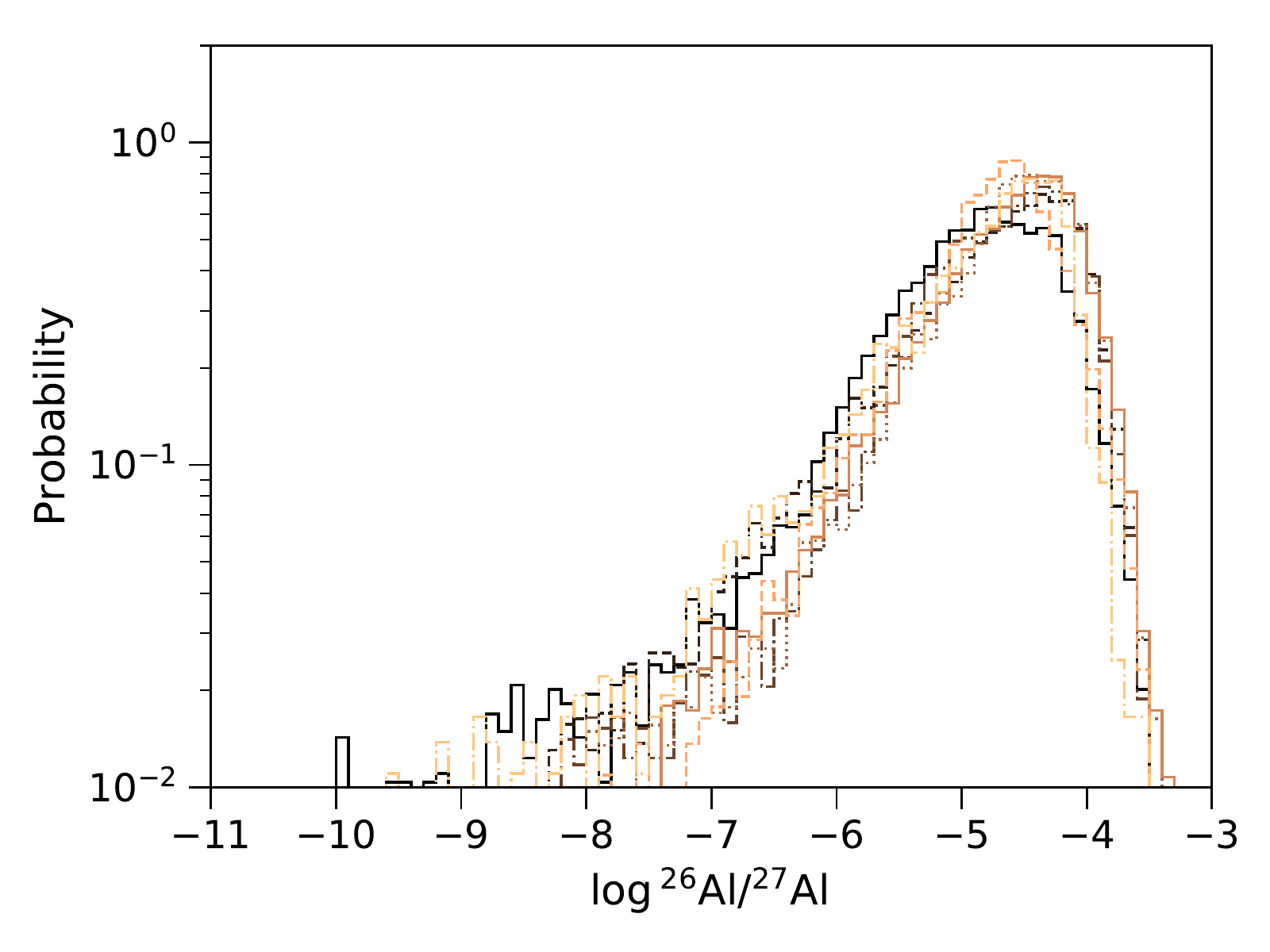}
\caption{Dependence of the $\rm ^{60}Fe/^{56}Fe$ and $\rm ^{26}Al/^{27}Al$ PDFs on Galactocentric radius. The histograms are each measured for stars formed within a 1 kpc-wide annulus centred at the Galactocentric radius indicated in the legend, at $t = 740-750$ Myr.}
\label{fig:PDF_abundance_ratio_radial}
\end{figure*}

To determine whether our stellar abundance distributions are typical in the whole galaxy, we examine the distributions for stars formed in 1 kpc-wide annuli centred on galactocentric radii from $4-10$ kpc. We show these radially-resolved distributions in \autoref{fig:PDF_abundance_ratio_radial}. The distributions clearly do not strongly vary with galactocentric radius. That means that most planetary system in the Galaxy could come to have the high abundance ratios of $\rm ^{60}Fe/^{56}Fe$ and $\rm ^{26}Al/^{27}Al$, and therefore the birth environment of the Solar System is not atypical not only near the Solar Circle but also for a broad region in the Milky-Way.

\begin{figure*}
\includegraphics[width=\columnwidth]{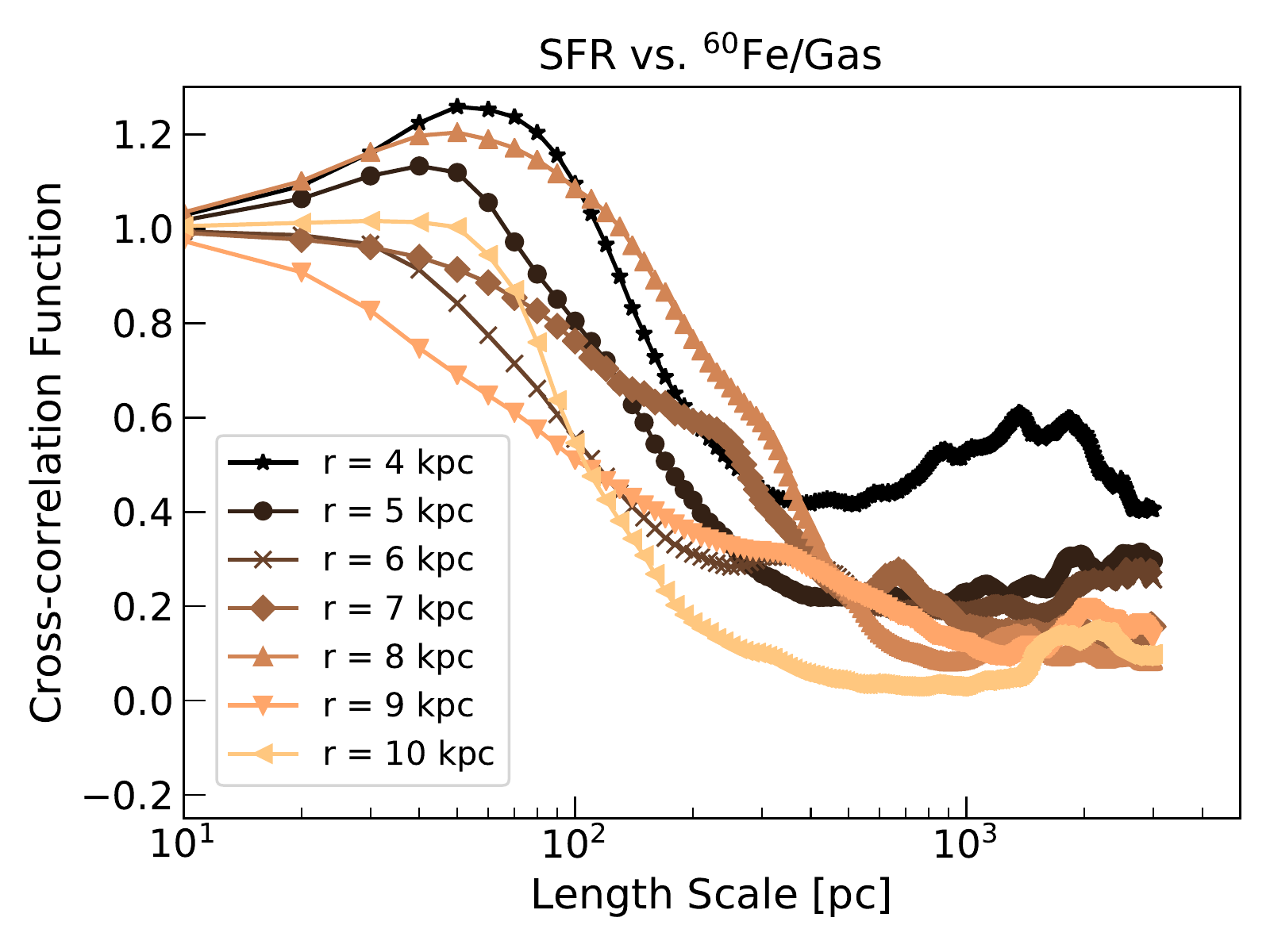}
\includegraphics[width=\columnwidth]{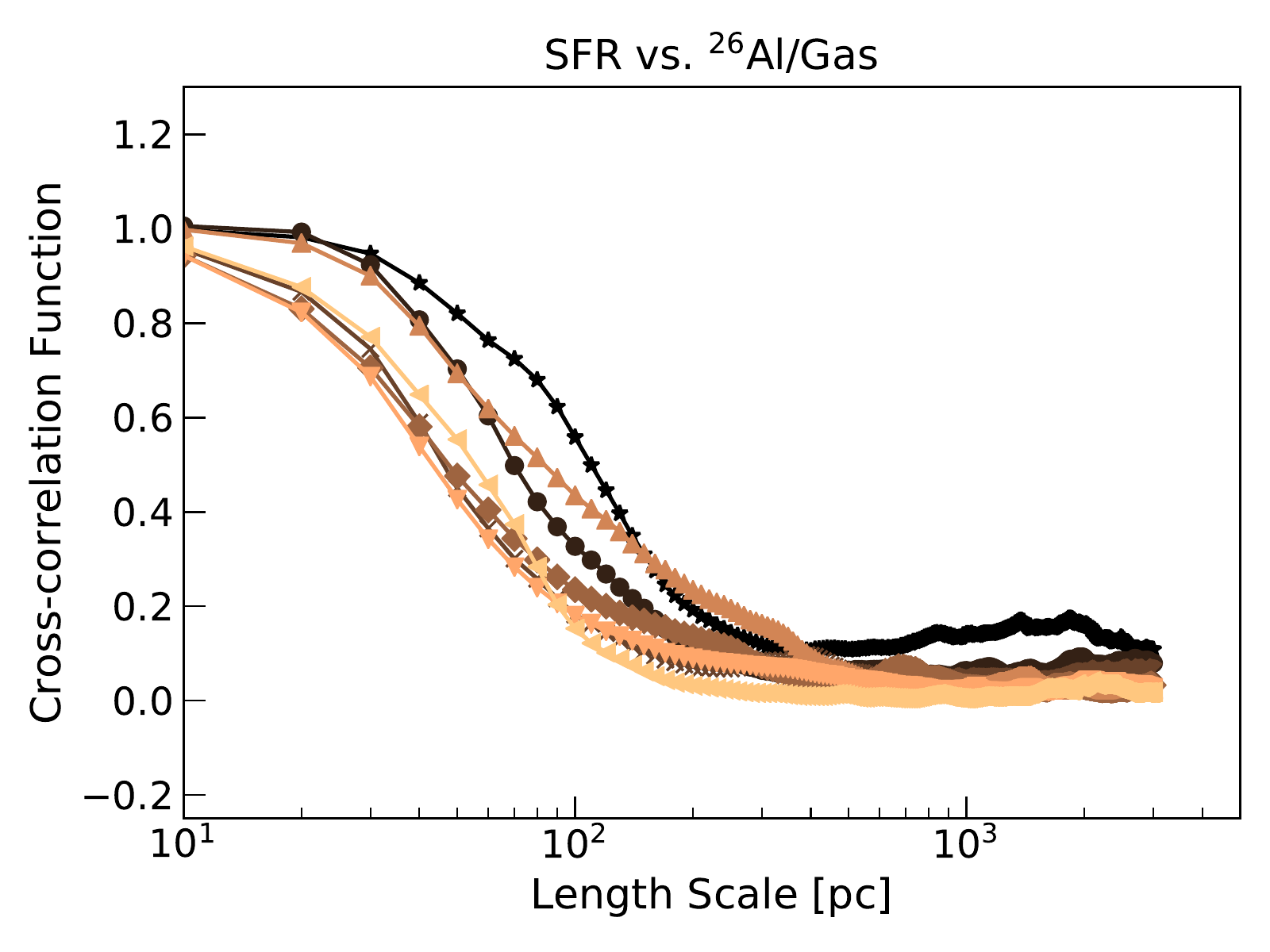}
\caption{Dependence of spatial cross-correlation functions between star formation and element abundance ratios on Galactocentric radius.}
\label{fig:cross_correlation_radial}
\end{figure*}

To determine if the physical explanation for these PDFs is the same at all galactocentric radii, in  \autoref{fig:cross_correlation_radial} we show the spatial correlation (\autoref{eq:correlation_function}) between star formation and SLR abundance measured at different galactocentric radii; we compute these functions using the same procedure as described in \autoref{Discussion}, except that we set the values of all pixels outside the target annulus to zero, so they do not contribute to the correlation. Although there is clearly some scatter in correlation with radius, the qualitative result that $^{60}\mathrm{Fe}$ correlates with star formation on scales of several hundred pc, and $^{26}\mathrm{Al}$ on scales of $\sim 100$ pc, appears to be the same at all galactocentric radii. This strongly suggests that the correlation is a result of the physics of stellar feedback and the lifetimes of the SLRs, rather than on any particular characteristic of the star-forming environment.

\section{Resolution and convergence}
\label{Resolution}

Because we find that the clustering of star formation is crucial to our results, it is important to demonstrate that the amount of clustering in our simulations is not artificially enhanced by our choice of star particle mass, since by construction stars that form within a single star particle are perfectly correlated. To investigate this possibility, we must verify that the true clustering scale of star formation in our simulations is much larger than the size of a single one of our star particles. We therefore calculate the two-point correlation function of star particles with ages $<1$ Myr, $\xi (r)$, which traces the amplitude of clustering of star particles as a function of scale. We perform this calculation using the clustering estimator of \citet{DavisPeebles1983},
\begin{equation}
\xi(r) = \frac{n_R}{n_D} \frac{DD}{DR} - 1.
\end{equation}
Here $DD$ is the number of star particle pairs with a separation in the range $r \pm \Delta r$ ($\Delta r = $ 5 pc) computed using the positions of stars output by our simulations (i.e., the ``data" catalogue, $D$), while $DR$ is the same quantity computed using pairs of particles where one is drawn from the actual list of stars ($D$), and the other is drawn from a ``random'' catalogue ($R$) generated by randomly placing stars in the same volume as $D$; $n_D$ and $n_R$ are the mean number densities of star particles in the data and random catalogues, respectively. For the purposes of our computation, we take our data catalogue to be the set of all star particles younger that 1 Myr at our final output time within a cubical region 2 kpc on a side, centred on the Solar circle; the region we use is the same one shown in \autoref{fig:density_projection_zoom}. For our random catalogue we use 1000 times as many random star particles as in the data catalogue.

\begin{figure}
\includegraphics[width=\columnwidth]{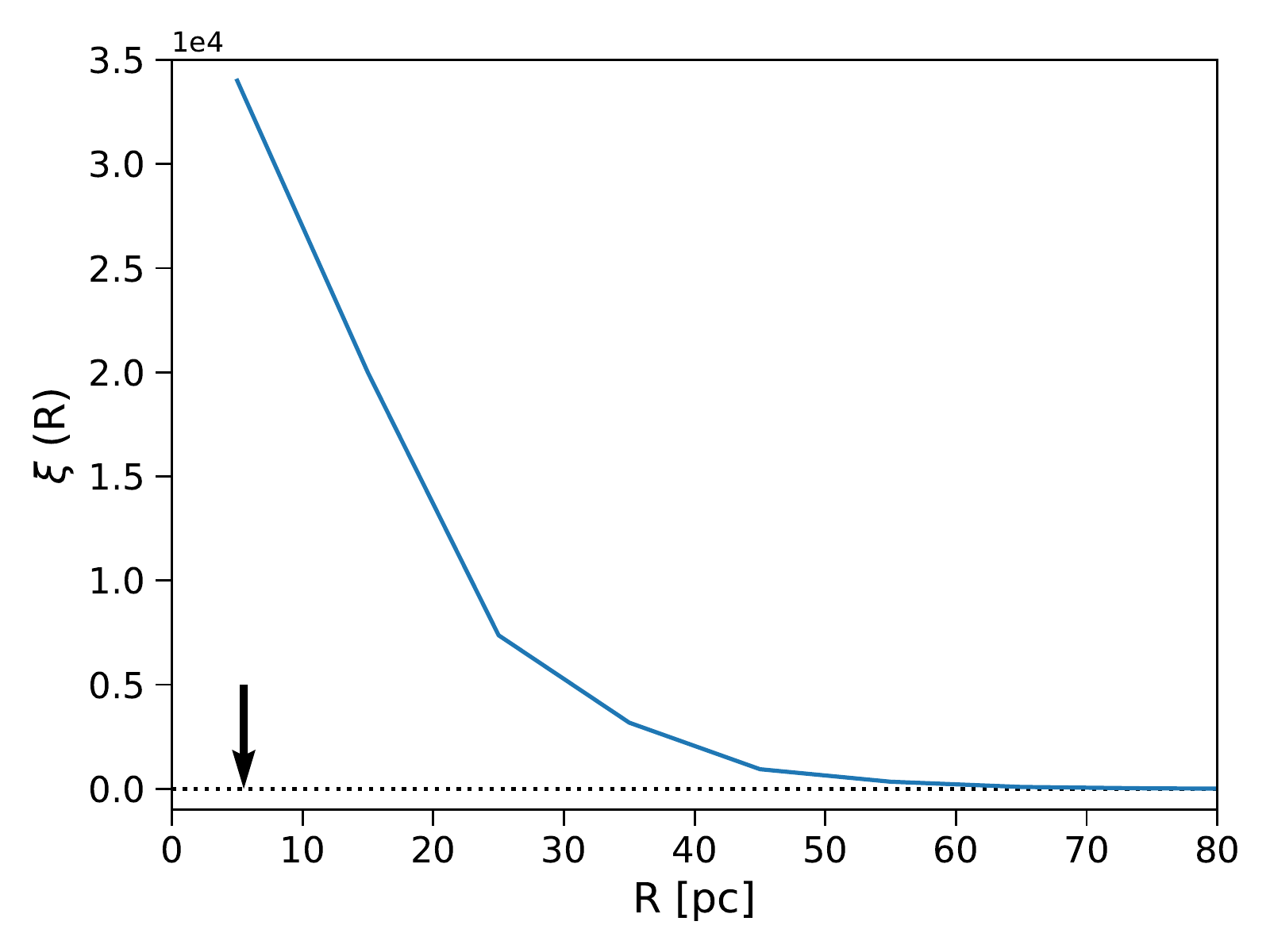}
\caption{Two-point correlation function of star particles, $\xi (R)$, which traces the amplitude of clustering of star particles as a function of scale. The arrow shows the size scale of star particle, defined as $(m_{\mathrm{sf}}/\rho_{\mathrm{sf}})^{1/3}$, where $m_{\mathrm{sf}}$ ($= 300\ M_{\odot}$) is star particle mass and $\rho_{\mathrm{sf}}$ ($= 57.5 m_p\ \mathrm{cm}^{-3}$) is the threshold density for star formation.}
\label{fig:two_point_correlation_particle}
\end{figure}

We show the result of this computation in \autoref{fig:two_point_correlation_particle}. We can see from the figure that the characteristic size scale on which star formation is clustered in our simulations is $\approx 40-50$ pc. For comparison, the size scale of ISM sampled by an individual star particle is  $\lesssim (m_{\mathrm{sf}}/\rho_{\mathrm{sf}})^{1/3}$, where $m_{\mathrm{sf}}$ ($= 300\ M_{\odot}$) is star particle mass and $\rho_{\mathrm{sf}}$ ($= 57.5m_p\ \mathrm{cm}^{-3}$ for mean particle mass $m_p$) is the threshold density for star formation; this is an inequality because gas does not form stars immediately upon reaching $\rho_{\rm sf}$, but may in fact collapse to somewhat higher density and smaller size scale before doing so. Our upper limit on the characteristic size of a star particle is 5.5 pc, which is shown in \autoref{fig:two_point_correlation_particle} as an arrow. Thus the size scales of stellar clustering in our simulation are roughly an order of magnitude larger than the sizes of individual star particles, and thus the choice of star particle size does not influence the degree of clustering.

\begin{figure*}
\includegraphics[width=\columnwidth]{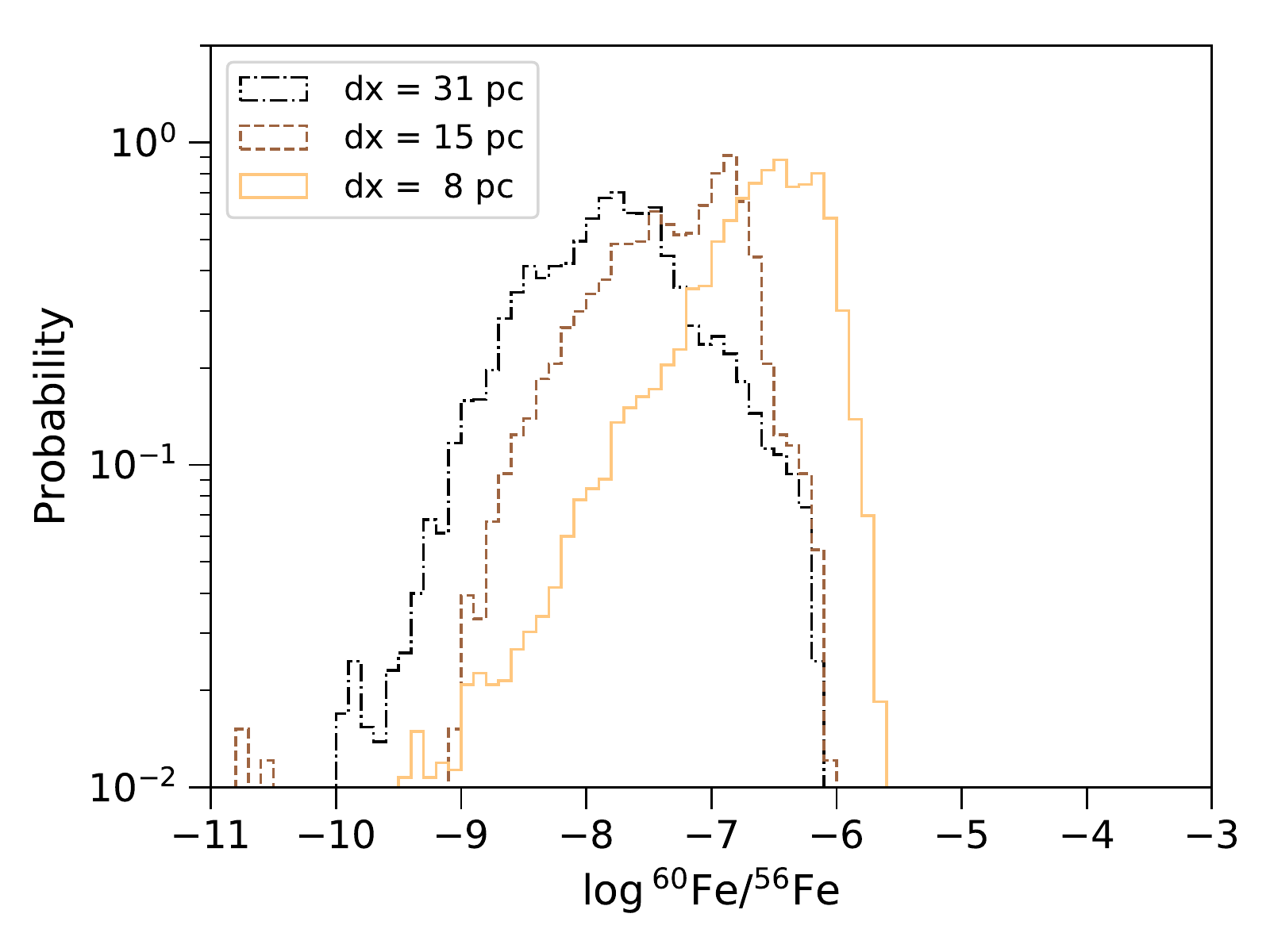}
\includegraphics[width=\columnwidth]{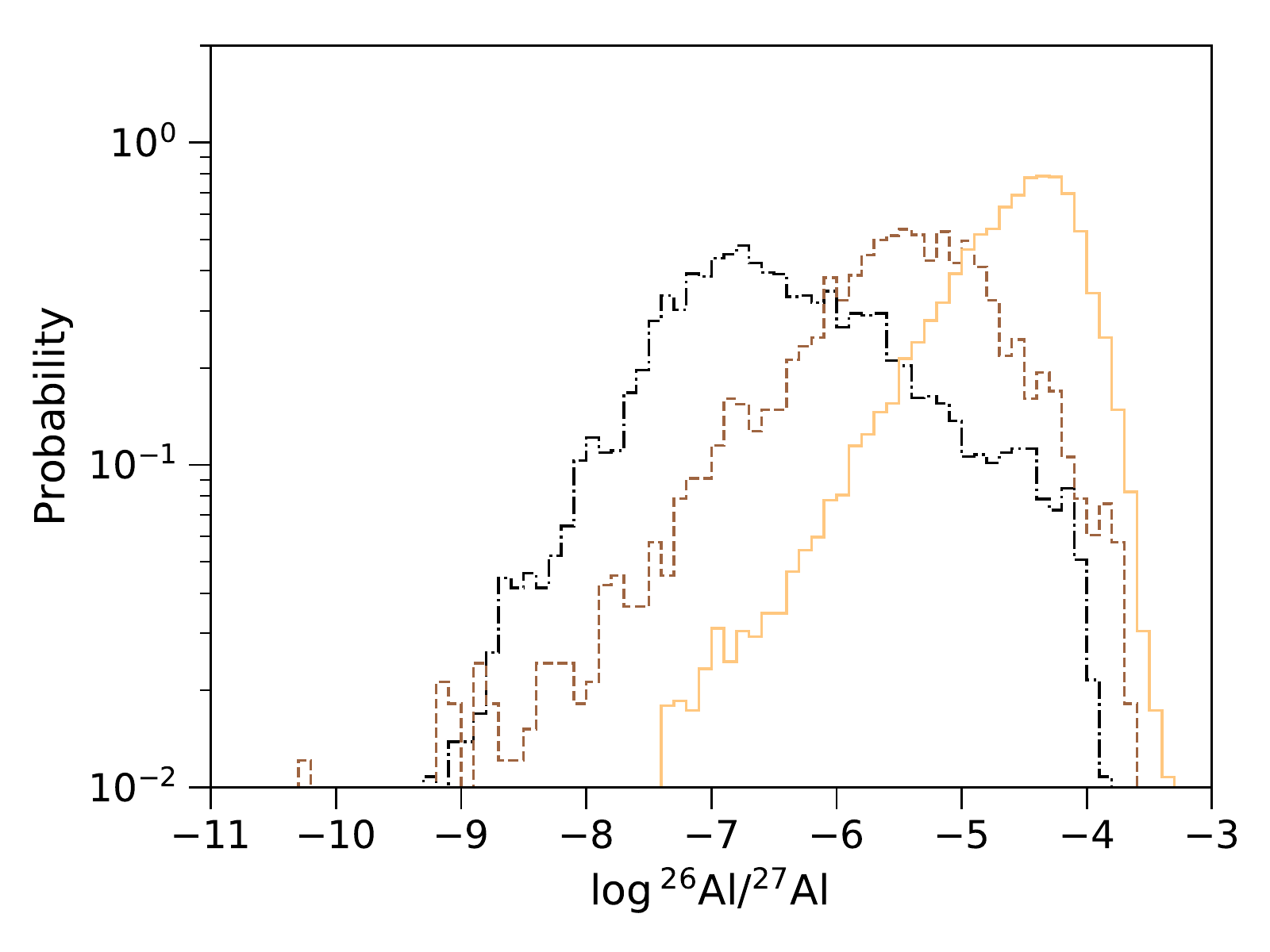}
\caption{Resolution study for SLR abundance PDFs in newly-formed stars at Galactocentric radii from 7.5 - 8.5 kpc. The left panel is $\rm ^{60}Fe/^{56}Fe$, and right is for $\rm ^{26}Al/^{27}Al$. The black dashdotted line shows the distribution of abundances at 31 pc resolution run ($t = 590-600$ Myr), the brown dashed line is 15 pc resolution ($t =650-660$) Myr, and the tan solid line is at 8 pc resolution ($t = 740-750$ Myr). }
\label{fig:PDF_abundance_ratio_resolution}
\end{figure*}

To determine whether our stellar abundance distributions are converged, we compare the distributions we measure for stars formed at $740-750$ Myr of evolution, when our resolution is 8 pc at the galaxy has reached steady state, to those formed at $590-600$ Myr (steady state at 31 pc resolution) and $650-660$ Myr (steady state at 15 pc resolution). We show the results in \autoref{fig:PDF_abundance_ratio_resolution}. We find that, although the peaks of the PDFs move to higher values with higher resolution, the high end tails converge to $10^{-6} \sim 10^{-5}$ for $\rm ^{60}Fe/^{56}Fe$, and $10^{-4} \sim 10^{-3}$ for $\rm ^{26}Al/^{27}Al$. Thus we are well-converged on the upper half of the abundance distribution. Moreover, given the broad range of uncertainties in the meteoritic abundance, the shifts we do see with resolution do not change the qualitative conclusion that Sun's SLR abundances are within the normal range expected for Milky-Way stars.

\bsp	
\label{lastpage}
\end{document}